\documentclass{article}
\usepackage{arxiv}
\usepackage{natbib}
\usepackage[utf8]{inputenc}             
\usepackage{algorithm}
\usepackage{slashbox}
\usepackage{graphicx}
\usepackage{color}
\usepackage[table]{xcolor}
\usepackage{collcell}
\usepackage{hhline}
\usepackage{pgf}
\usepackage{amssymb,amsmath}
\usepackage{amsfonts}
\usepackage{algorithm,algorithmic}
\usepackage[table]{xcolor}
\usepackage{algorithmic}
\usepackage[export]{adjustbox}
\usepackage[small]{caption}
\usepackage{mwe} 
\usepackage{lipsum}
\usepackage{array}
\usepackage{subfig}
\usepackage{gensymb}
\usepackage{comment}
\usepackage{epstopdf}
\usepackage{multirow}
\graphicspath{ {Final_Figs/} }

\def\colorModel{hsb} 
\newcommand\ColCell[1]{
	\pgfmathparse{#1<0.5?1:0}  
	\ifnum\pgfmathresult=0\relax\color{white}\fi
	\pgfmathsetmacro\compA{0}      
	\pgfmathsetmacro\compB{#1/0.8} 
	\pgfmathsetmacro\compC{1/0.5}      
	\edef\x{\noexpand\centering\noexpand\cellcolor[\colorModel]{\compA,\compB,\compC}}\x #1
} 
\newcolumntype{E}{>{\collectcell\ColCell}m{0.7cm}<{\endcollectcell}}  

	\title{Asynchronous Averaging of Gait Cycles for Classification of Gait and Device Modes}
  	\author{Parinaz Kasebzadeh \\  Department of Electrical Engineering\\
	Link\"oping University\\Link\"oping, Sweden\\ \texttt{parinaz.kasebzadeh@liu.se} 
	\And Gustaf Hendeby \\ Department of Electrical Engineering\\
	Link\"oping University\\Link\"oping, sweden\\\texttt{gustaf.hendeby@liu.se}  \And Fredrik Gustafsson\\  Department of Electrical Engineering\\
	Link\"oping University\\Link\"oping, sweden\\\texttt{fredrik.gustafsson@liu.se} }
\begin{document}
	
	\maketitle
	
	\begin{abstract}
	An approach for computing unique gait signature using measurements collected from body-worn inertial measurement units (IMUs) is proposed. The gait signature  represents one full cycle of the human gait, and is suitable for off-line or on-line classification of the gait mode. The signature can also be used to jointly classify the gait mode and the device mode. The device mode identifies how the IMU-equipped device is being carried by the user. The method is based on precise segmentation and resampling of the measured IMU signal, as an initial step, further tuned by minimizing the variability of the obtained signature within each gait cycle. Finally, a Fourier series expansion of the gait signature is introduced which provides a low-dimensional feature vector well suited for classification purposes. 
	The proposed method is evaluated on a large dataset involving several subjects, each one containing two different gait modes and four different device modes. The gait signatures enable a high classification rate  for each step cycle.
	\end{abstract}
	
		\keywords {Pedestrian dead reckoning, gait cycles, inertial measurement unit (IMU)}


\section{Introduction}
Today, a variety of gadgets are attached  to our bodies that contain an inertial measurement unit (IMU), for instance smartphones, smart watches, VR headsets, cameras, etc. There are also tailored solutions like foot-mounted IMUs for firefighters and IMU-equipped body suits used by e.g. the movie and gaming industry for virtual reality motion capture~\cite{journal:Chowdhury2018}. 
Besides solving dedicated tasks, these IMU signals also contain what will be referred to as the {\em gait signature} that is caused by the steps we take when moving. 
Examples include bio-mechanical analysis of limping patterns for diagnosis of certain deceases such as Parkinson's disease~\cite{journal:Caramia2018}, and dead-reckoning for indoor positioning systems \cite{Journal:Ruiz2012,journal:Tian2014,Journal:Davidson2017,Journal:Jimenez2014}. We will refer to such classification tasks as determining the {\em gait mode}.
The gait mode can also include bio-mechanical hypotheses of anomalies.
 
In most such studies, the position of the IMU on the body is known and decided by the application. Applications include foot-mounted IMUs for firefighters, head-mounted IMUs for VR glasses, wrist-mounted IMUs for smart watches, and IMUs for various fixed body parts in body suits~\cite{Symp:Nilsson2014,Journal:Alvarez2012, Journal:Skog2010,Journal:Zeng2018,Journal:Panahandeh2013}. For more general applications involving for example smartphones, the position might vary over time. We will refer to the position of the sensor on the body as the {\em device mode}. 

There is a rich literature on the subject of using IMU signals for pedestrian dead reckoning (PDR)~\cite{journal:Basso2017, Balachandran2003, Journal:Ho2016, Journal:Zihajehzadeh2015, journal:Diez2018}. PDR is one application of gait analysis, which is typically solved by thresholding techniques, where e.g. the norm of the accelerometer is first band-pass filtered and then thresholded~\cite{Journal:Brzostowski2018, journal:Norrdine2016, Conf:Kasebzadeh2016}. Using a fixed threshold typically leads to systematic errors for people heavier or lighter than the test subjects the threshold is designed for, and there are also false positive and negative step detections on the test subjects themselves. 
A third application of the gait signature is to improve step detection. 

 Step detection in PDR normally relies on zero velocity update (ZUPT) for lower-body mounted IMUs, such as foot-mounted applications. ZUPT is a typical,  self-contained technique for step detection  that benefits from the cyclical nature of human walking patterns and eliminates the bias from accelerometer and gyroscope~\cite{Journal:Zeng2018, Journal:Skog2010, journal:Norrdine2016}. 
However, extra care must be taken when dealing with upper-body sensors. The upper-body mounted or hand-held IMUs might report continuous or unexpected  motion while the sensors in the lower body capture the foot at rest. Hence, instead of finding zero velocity periods as in ZUPT, step detection in PDR normally relies on peak detection or  threshold-based approaches~\cite{Jornal:Renaudin2012, Journal:Kang2015}.  

Accurate step detection, hence, requires joint gait mode and device mode classification in order to get the proper threshold for the peak detection~\cite{Journal:Pappas2001, Journal:Ho2016, journal:Zhang2015,Journal:Zhang2018}.  
An adaptive gait detection and step length estimation, based on walking speed classification, is proposed in~\cite{Journal:Ho2016}.  
A probabilistic, user-independent method, as an alternative to threshold-based approaches, is introduced in~\cite{Journal:Panahandeh2013} that uses chest-mounted IMUs to jointly perform gait analysis and classify the activity motions.  A weighted context-based step length estimation  algorithm using waist-mounted IMUs embedded in smartphones is proposed in~\cite{Journal:Martinelli2018} which strives to classify six different pedestrian activities. 

Existing approaches for step detection and gait cycle segmentation typically rely on  measurements collected from hand-held devices such as smartphones that are already equipped with IMUs. Using such devices does not impose any extra cost and can become a universal solution~\cite{Journal:Li2016,Journal:Kang2015}.  However, due to the large number of factors affecting  the sensor readings, such as the user’s motion mode and the device  mode, these methods suffer from robustness issues and might collapse if the underlying assumption is not satisfied. One solution to the problem is to classify the mode of the system and use the additional information obtained from this knowledge to robustify the algorithm.

The gait signature as observed by the IMU depends on both the gait mode (e.g. running, walking, strolling) and the device mode (for instance, a smartphone can be held in the hand, stored in a pocket or backpack, etc.), and as such reveals a rich information source suitable for a variety of applications. Our key contribution is a proposed algorithm for off-line analysis of IMU data during motion, with the following outline:
\begin{enumerate}
\item Gait segmentation using optimization to maximize similarity of the gait cycles. This step might need initialization, and here classical step detection algorithms can be used.
\item Estimation of the gait signature by averaging over the segments. This is done on a normalized time scale, so small variations in step cycle times are handled by resampling techniques. 
\item Extraction of a low dimensional feature vector for the gait cycle using Fourier series analysis on the estimated gait signature.  This feature vector includes physically explainable patterns. 
\end{enumerate}
This algorithm to estimate the gait signature is presented as an off-line one, although on-line versions are plausible. 
Thus, the gait signature estimation method can be used for either on-line classification, or off-line gait analysis.

The remainder of this paper is organized as follows: 
Sec.~\ref{Sec:PF} describes the considered problem in detail. In Sec.~\ref{sec:Opt}, a standard threshold-based method for step detection is given followed by the solution to the gait cycle optimal segmentation problem. Sec.~\ref{sec:FS} presents a  Fourier series approach to achieve a low-dimensional feature vector. The performance of the proposed method is evaluated in Sec.~\ref{sec:results}, which also includes an optional data pre-processing stage. Finally, the work is concluded in Sec.~\ref{sec:con}.
\section{Problem Formulation and Notation}
\label{Sec:PF}
The most frequently used notations in this paper are summarized in Table~\ref{notation}.
\begin{table}[htp]
\caption{Notation}
\begin{center}
\begin{tabular}{l|l}
$\boldsymbol{x}(s)$ & Noise-free data \\
$\boldsymbol{y}(s)$ & Measured data \\
$\boldsymbol{e}(s)$ & IMU measurement noise \\
$\boldsymbol{\hat g}_m(\tau)$ & $m$:th gait cycle  \\
$\boldsymbol{\bar g}(\tau)$ & Gait signature  \\
$\boldsymbol{\hat G}[l]$ & Fourier series expansion of $\boldsymbol{\bar g}(\tau)$  \\
$N$ & Number of IMU samples\\
$M$ & Number of gait cycles\\
$L$ & Number of grid points of normalized time\\
$s_n$ & Sample times of IMU \\
$t_m$ & Step time for $m$:th gait cycle \\
$\tau$ & Normalized time $\tau\in[0,1)$ \\
$\epsilon_{\mathrm{lo}}$ & Minimum gait cycle time \\
$\epsilon_{\mathrm{up}}$ & Maximum gait cycle time\\
$\epsilon_p$ & Peak threshold \\
$\epsilon_v$ & Valley threshold \\
\end{tabular}
\end{center}
\label{notation}
\end{table}%

\begin{figure}[!t] 
	\centering
	\subfloat[Gait cycles, marked with blue lines, and averaged gait signature, marked with the green line.
\label{manualGCa}]{\includegraphics[width=0.5\textwidth]{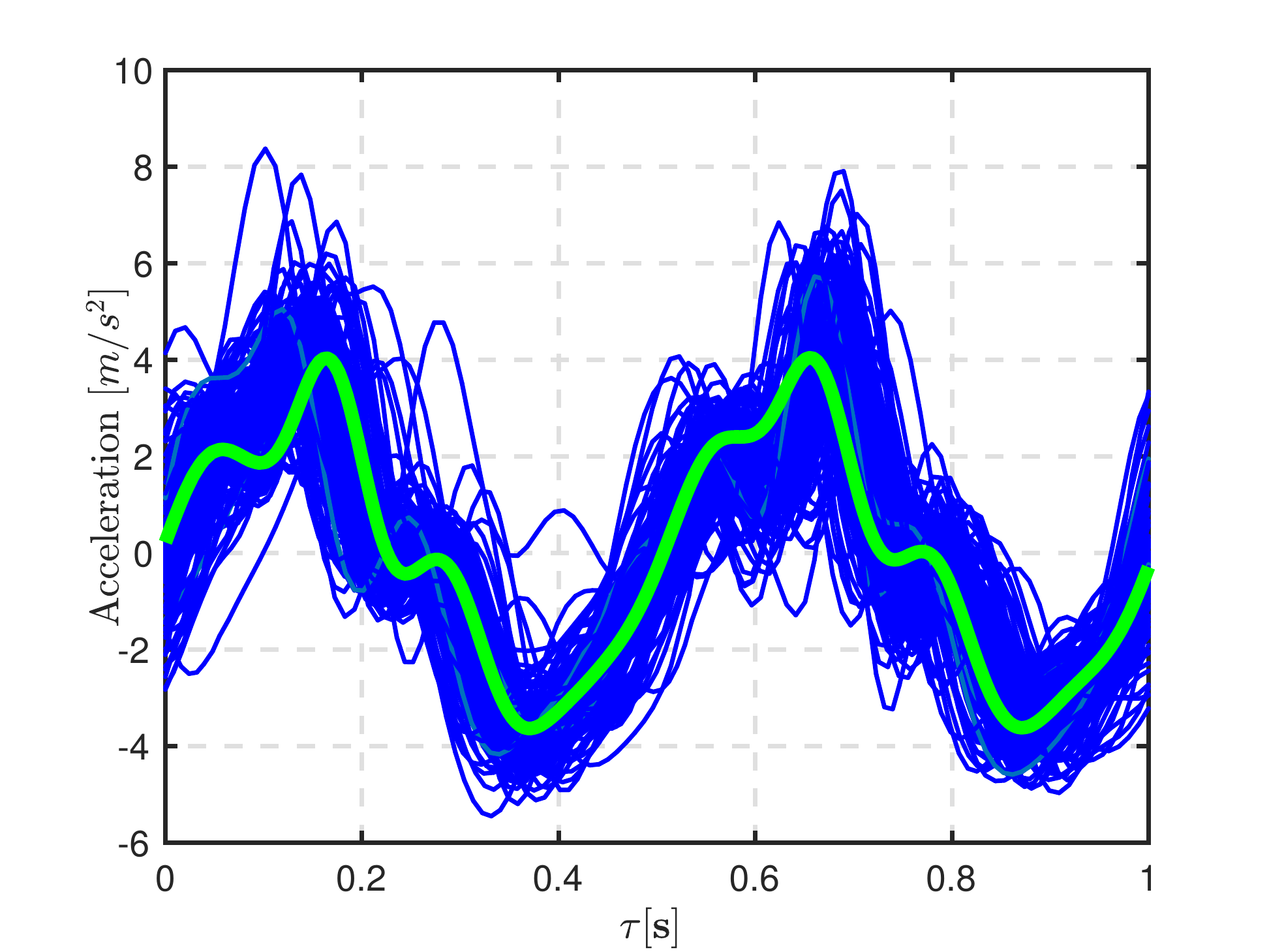}}
	\subfloat[Histogram of step times. \label{manualGCb}]{\includegraphics[width=0.5\textwidth]{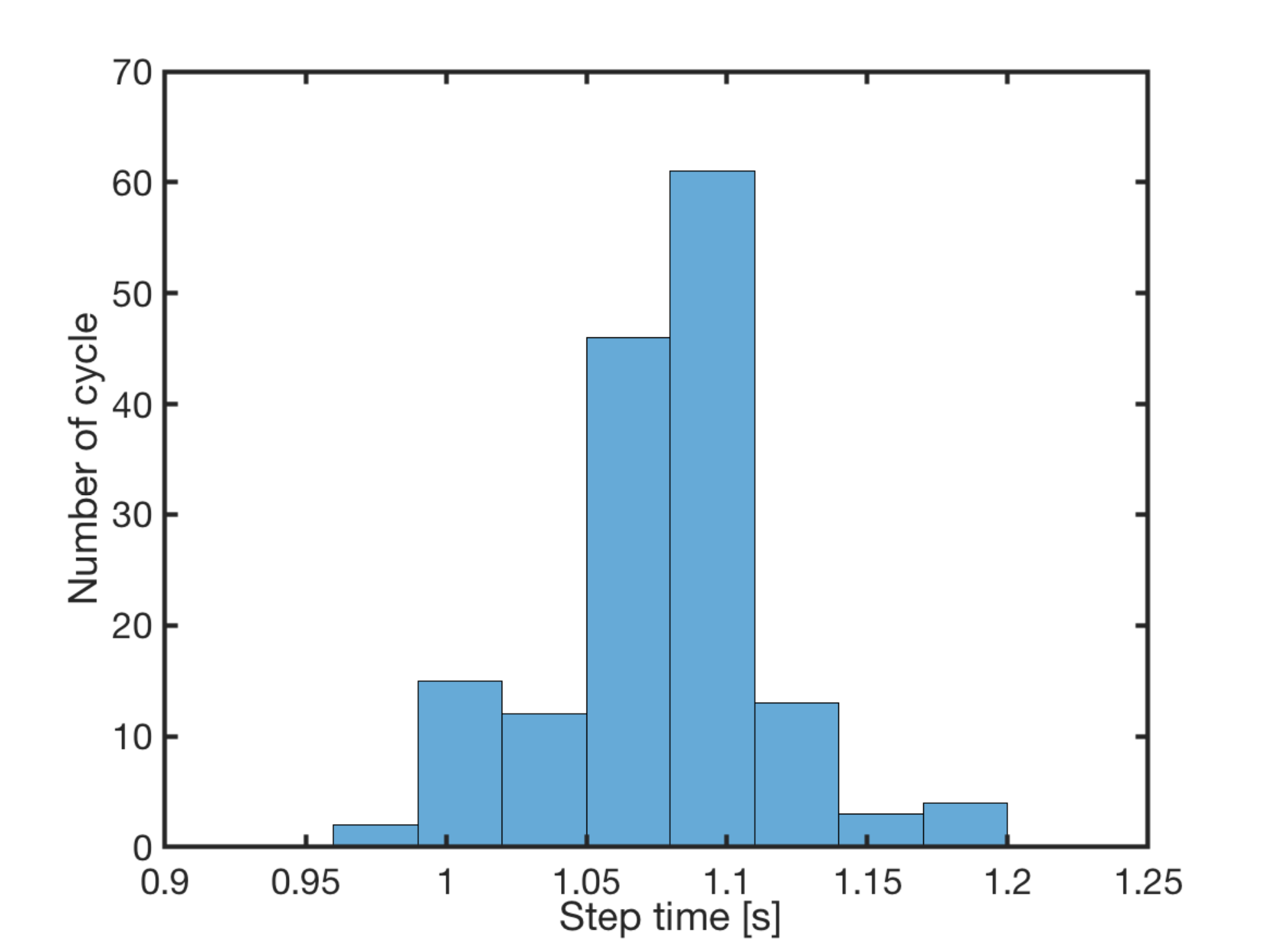}}
	\caption{Gait cycles, manually segmented using accelerometer signal.}
	\label{manualGC}
\end{figure}

Consider some physical quantity $\boldsymbol{x}(s)$, possibly multi-dimensional,  that depends on gait, which can be measured with additive noise $\boldsymbol{e}(s)$ as
\begin{align}
\boldsymbol{y}(s_n) = \boldsymbol{x}(s_n)+\boldsymbol{e}(s_n), \quad n=1,2,\dots,N,
\end{align}
where $s_n$ denotes the sampling times.  
We assume that the gait cycle is periodic over periods of time, and we are interested in the underlying average gait cycle of the physical quantity $\boldsymbol{g}(\tau)$ on a normalized time scale $0\leq \tau < 1$. Fig.~\ref{manualGCa} is an example to represent periodic gait cycles over the normalized time scale. The presented data in Fig.~\ref{manualGCa} corresponds to  a real scenario in which the walking subject was carrying a smartphone in the hand, facing upwards. All the gait segments in the figure were extracted manually from a bandpass-filtered accelerometer signal. 

The challenge is to estimate the gait cycle $\boldsymbol{\hat g}(\tau)$ from the IMU measurements $\boldsymbol{y}(s)$, and the most critical step is the segmentation in which the signal is split into the separate gait cycles. 
 We denote the beginning of each gait cycle time by  ${t}_m$ such that 
\begin{align}
s_1 \leq {t}_0 < {t}_1<\dots < {t}_M \leq s_N.
\end{align}
Fig.~\ref{manualGCb} shows a histogram of step time duration, $t_m-t_{m-1}$, for all $M$ gait cycles in Fig.~\ref{manualGCa}.
Given these durations, we can obtain a estimate of the gait cycle, Fig.~\ref{manualGCa}, from each segment as
\begin{align}
\boldsymbol{\hat{g}}_m(\tau) = x({t}_{m-1} + ({t}_{m}-{t}_{m-1})\tau), \quad \tau\in[0, 1).
\end{align}
Then, we immediately get what is referred to as an {\em asynchronously averaged} gait cycle by
\begin{align}
\boldsymbol{{\bar{g}}}(\tau) = \frac{1}{M} \sum_{m=1}^M \boldsymbol{\hat{g}}_m(\tau). 
\label{signature}
\end{align}
The green line in Fig.~\ref{manualGCa} represents $\boldsymbol{\bar{g}}(\tau)$ for all manually extracted gait cycles.

The key problem is thus to determine the step times such that different gait cycles $\boldsymbol{\hat{g}}_m(\tau) $ become as similar as possible. The considered metric as a measure of the similarity is the variance between each gait cycle and the averaged gait signature. We propose a nonlinear least squares framework where we optimize the step times in order to minimize the variance of these gait cycles
\begin{subequations}
\begin{equation}
\label{eq:genopt}
\hat{t}_{0{:}M} = \arg\min_{{t}_{0{:}M}}  V({t}_0{:}{t}_M),
\end{equation}
\begin{equation}
\label{eq:genopt2}
V({t}_0{:}{t}_M)= \frac{1}{M} \sum_{m=1}^M  \left\|  \boldsymbol{\bar{g}}(\tau) - \boldsymbol{\hat{g}}_m(\tau) \right\| ^2.
\end{equation}
\end{subequations}

To make the optimization problem mathematically tractable, we will approximate the optimization problem in the following ways:
\begin{enumerate}
\item The normalized time scale is discretized to $L$ uniformly spaced grid points $\tau_l=(l-1)/L$ for $l=1,2,\dots,L$. 
\item The $L_2$-norm is used in \eqref{eq:genopt2}, which gives a nonlinear least squares (NLS) problem, which often allows for efficient solvers.
\end{enumerate}

Based on these assumptions, the problem can be written as
\begin{align}
\label{eq:genopt3}
\hat{{t}}_{0{:}M} 
&= \arg\min_{{t}_{0{:}M}}  \frac{1}{LM} \sum_{m=0}^M  \sum_{l=1}^L \left( \boldsymbol{ \bar{g}}(\tau_l) - \boldsymbol{\hat{g}}_m(\tau_l) \right)^2. 
\end{align}
%

These two restrictions also enable us to formulate the signal Fourier series (FS) expansion  in order to reduce the model order as  introduced  in Sec.~\ref{sec:FS}.

A solution for this optimization problem is proposed  in Sec.~\ref{sec:optimization}. Moreover, 
we will use a rather standard step detection method based on thresholding, outlined in Sec.~\ref{sec:GD}, in order to initialize the optimization problem. 

\section{Optimal Segmentation of Gait Cycles}
\label{sec:Opt}
%
In this section, we first suggest a solution to the optimization problem~\eqref{eq:genopt3}. Then,  a standard threshold-based step detection method is presented in order to initialize the optimization algorithm. 

It is worth noting that all solutions are provided based on the assumption that the measurements are collected in advance (the solution is for the off-line mode). However, it could be easily extended to the on-line problems.  Moreover, during the optimization process, all the gait cycles should be in the same regime of gait and device modes introduced in Table~\ref{tbl:classes}.
\subsection{Solution for Optimization Problem}
\label{sec:optimization}
The sensor data should be band-pass filtered before segmentation to remove slow trends and high frequency noise. The cut-off limits in the band-pass filter should be selected to take the slowest and fastest pace into account --see Sec.~\ref{sec:GD} for further details. Moreover,  the minimum step time $\epsilon_{lo}$ and the maximum step time $\epsilon_{up}$ are required to be defined in advance. 
These bounds can be obtained from the most frequent time interval of the gait cycles given by the histogram, {\em e.g.} Fig.~\ref{manualGCb}, of the detected gait cycles. The estimated step cycles should be in this interval. 

The  minimization problem for the objective function $V({t}_0{:}{t}_M)$ is given by
%
%
\begin{equation}
\begin{aligned}
& \underset{{t}_0{:}{t}_M}{\text{minimize}}
& & V({t}_0{:}{t}_M)\\
& \text{subject to}
&& {\epsilon_{lo}<{t}_m-{t}_{m-1}<\epsilon_{up}} \\
\end{aligned}
\label{basicOptimization}
\end{equation}
%
%

In order to make the optimization procedure feasible, it is reformulated into $M$ sub-optimal problems where each problem requires the optimal solution from the previous one. The first sub-problem is defined by 
	\begin{equation}
	\begin{aligned}
	& \underset{{t}_1}{\text{minimize}}
	& & V({t}_0{:}{t}_M), \\
	& \text{subject to}
	&& { \epsilon_{lo}<{t}_1-{t}_{0}<\epsilon_{up}}\\
	\end{aligned}
	\label{eq:reformulated_step_1}
	\end{equation}
where the outcome of the problem would be the optimal value of $t_1$  denoted by $\hat t_1$. Given $\hat t_1$, the second sub-problem becomes
\begin{equation}
\begin{aligned}
& \underset{{t}_2}{\text{minimize}}
& & V({t}_0,\hat{t}_1,t_2{:}{t}_M), \\
& \text{subject to}
&& { \epsilon_{lo}<{t}_2-\hat{t}_{1}<\epsilon_{up}}\\
\end{aligned}
\label{eq:reformulated_step_2}
\end{equation}
The optimized $\hat t_2$  is the outcome of the second step. 

To generalize, the $i{:}{\mathrm{th}}$ $i\in[2,\ldots,M]$  sub-problem is given by
	\begin{equation}
	\begin{aligned}
	& \underset{{t}_i}{\text{minimize}}
	& & V(t_0,\hat{t}_1{:}\hat{{t}}_{i-1},{t}_i{:}t_M).\\
	& \text{subject to}
	&& { \epsilon_{lo}<{t}_i-\hat{t}_{i-1}<\epsilon_{up}}\\
	\end{aligned}
	\label{eq:reformulated_step_i}
	\end{equation}
These simplified optimization problems could be solved using a general linear search method algorithm which iteratively minimizes the cost function considering the given boundaries. Derivative-free quadratic interpolation and golden section search methods are used in this work~\cite{Book:Nocedal2006}. 

%
%
%
%
%
\begin{algorithm}[!t]
	\caption{Proposed gait cycle segmentation algorithm}
	\begin{algorithmic}[1]
		\renewcommand{\algorithmicrequire}{\textbf{Input:}}
		\renewcommand{\algorithmicensure}{\textbf{Output:}}
		\REQUIRE $\hat g(\tau)$, ${\boldsymbol t}=\{t_0{:}t_M\}$, $\epsilon_{lo}$,  $\epsilon_{up}$ and $\gamma$.
		\ENSURE $\boldsymbol {\hat t}=\{\hat t_1{:}\hat t_M\}$.
		\STATE Initialization:\\
		- Compute $\boldsymbol{{\bar{g}}}(\tau)$ using~\eqref{signature}  \\
		- Compute ${V_{1}(t_0{:}t_M)}$ using~\eqref{eq:genopt3}  \\
		- set  $i=2$\\
		\REPEAT
		\STATE  set $m=1$.
		\WHILE{$m<=M$}
		\STATE Find $\hat t_m$ using~\eqref{eq:reformulated_step_i} for $t_m$
		\STATE Replace $t_m=\hat t_m$
		\STATE Update $\boldsymbol{{\bar{g}}}(\tau)$ using~\eqref{signature} 
		\STATE $m=m+1$
		\ENDWHILE
		\STATE Compute $V_{i}(t_0{:}t_M)$ using~\eqref{eq:genopt3}  
		\IF {$V_{i}(t_0{:}t_M)-V_{i-1}(t_0{:}t_M) < \gamma$ }
		\STATE $\boldsymbol {\hat t}=\{\hat t_1{:}\hat t_M\}$
		\STATE Terminate the iterations.
		\ENDIF
		\STATE $i=i+1$
		\UNTIL iterations are terminated.	
	\end{algorithmic} 
	\label{Opt_Alg}
\end{algorithm}
%
%
%
%
Algorithm~\ref{Opt_Alg} outlines the proposed solution to find the optimal gait cycle segments from a given  accelerometer signal. To initialize the algorithm,  the preliminary gait cycles, $ \boldsymbol{\hat g}_{1{:}M}(\tau)$, are detected using a classical threshold-based step detection algorithm as described in Sec.~\ref{sec:GD}. The initial signature, $\boldsymbol{{\bar{g}}}(\tau)$, is then estimated  by~\eqref{signature}  considering all detected $\boldsymbol{\hat g}_{1{:}M}(\tau)$. Subsequently,  the cost function,  $V(t_0{:}t_M)$, introduced in~\eqref{eq:genopt2} is computed using $\boldsymbol{{\bar{g}}}(\tau)$ and  $\boldsymbol{\hat g}_{1{:}M}(\tau)$.
For each gait cycle, $\boldsymbol{\hat g}_i(\tau)$, the algorithm strives to find the optimal value $\hat t_i$ using~\eqref{eq:reformulated_step_i}.  Additionally, the initial gait signature is updated based on the new set of $\boldsymbol{t}$. 

Once all the detected gait cycles have been optimized, the cost function,  $V(t_0{:}t_M)$, will be re-computed given the estimated gait cycles. Henceforth, the new optimal gait cycles and updated signature will be used as the initialization for the next iteration of optimization problems. 
The termination criterion for this iterative algorithm is based on the decrement on the cost function.  Hence, the algorithm will be iterate until the absolute decrease of cost function falls below the given threshold $\gamma$. 

The performance of the optimization problem is evaluated for all introduced scenarios in Table~\ref{tbl:classes} in Sec.~\ref{sec:results}.  
%
%
\subsection{Classical Gait  Segmentation}
\label{sec:GD}
To initialize the algorithm for offline applications, a threshold based step detection method is employed.
A gait cycle, containing two consecutive toe-off moments of the same foot, can be crudely detected by defining two thresholds.  If the peak threshold $\epsilon_p$ and the valley threshold $\epsilon_v$ are each crossed twice, then a cycle is taken. 
 The  measured signal will  be compared against both $\epsilon_p$ and $\epsilon_v$ to find the time interval that it takes for the signal to cross both thresholds twice. Each gait cycle is then defined as sample values along this interval.

All the steps of the threshold-based step detection algorithm are  summarized in Algorithm.~\ref{GCDA}. As a first step, in the gait cycle detection algorithm, we need to define the hyperparameter values for $\epsilon_p$ and $\epsilon_v$. These values depend on human activity mode and device mode. Different approaches are introduced to consider proper thresholds, either  pre-defined or adaptive~\cite{journal:Zhang2015,Conf:Kasebzadeh2016,Patent:Levi1996}. In this work, pre-defined  thresholds on norm of acceleration value  have been used to detect peaks and valleys.

By considering a proper threshold for each scenario, the gait cycle can be extracted by comparing the norm of  acceleration signal at each time  with the $\epsilon_p$ or $\epsilon_v$. Then, the upper bound or lower bound is hit  as soon as  $||\boldsymbol{ a}(k)||$  becomes larger than $\epsilon_p$  or smaller than $\epsilon_v$, respectively.  Noting that each gait cycle contains two peaks and two valleys, $c_{\mathrm{up}}$ and $c_{\mathrm{lo}}$ are defined  in order to count the number of times that the signal, $||\boldsymbol{ a}(k)||$, hits the thresholds. 
%
%
%
%
%
\begin{algorithm}[!t]
	\caption{Gait cycle detection}
	\begin{algorithmic}[1]
		\renewcommand{\algorithmicrequire}{\textbf{Input:}}
		\renewcommand{\algorithmicensure}{\textbf{Output:}}
		\REQUIRE 
		Norm of accelerometer signal
		\begin{align}
	 ||{\boldsymbol{a}}(k)||=\sqrt{\boldsymbol{a}_x^2+\boldsymbol{a}_y^2+\boldsymbol{a}_z^2} .\nonumber
		\end{align}
		- Lower bound $\epsilon_{\mathrm{lo}}$ and upper bound $\epsilon_{\mathrm{up}}$ thresholds, if hit step is occurred
		\ENSURE Set of gait cycles $\{ Y_m\}_{m=1}^{M}$
		\STATE Initialization:\\
		- Set counters $c_{\mathrm{up}}=0$ and $c_{\mathrm{lo}}=0$ representing the number of times upper and lower bounds are hit\\
		- set $hit_{p}=\text{FALSE} $ and $hit_{v}=\text{FALSE} $\\
		- $k_{start}=1$\\
		- $k=k_{start}$\\
		- $m=1$
		\REPEAT 
		\IF{$||{\boldsymbol{a}}(k)||\geq\epsilon_{\mathrm{up}}$} \STATE$c_{\mathrm{up}}=c_{\mathrm{up}}+1$
		\STATE$hit_{p}=\text{TRUE} $, $hit_{v}=\text{FALSE} $
		\ENDIF
		\IF{$||{\boldsymbol{a}}(k)||\leq\epsilon_{\mathrm{lo}}$ } 
		\STATE$c_{\mathrm{lo}}=c_{\mathrm{lo}}+1$
		\STATE$hit_{p}=\text{FALSE} $, $hit_{v}=\text{TRUE} $
		\ENDIF
		\STATE $k=k+1$
		\UNTIL $hit_{p}=\text{TRUE} $ or $hit_{v}=\text{TRUE} $
		\WHILE{$c_{\mathrm{up}}<2$ and $c_{\mathrm{lo}}<2$ and $k\le N$}
		\IF{$||{\boldsymbol{a}}(k)||\geq\epsilon_{\mathrm{up}}$ and $hit_{p}=\text{FALSE} $ and $hit_{v}=\text{TRUE} $}
		 \STATE$c_{\mathrm{up}}=c_{\mathrm{up}}+1$
		\STATE$hit_{p}=\text{TRUE} $, $hit_{v}=\text{FALSE} $
		\ENDIF
		\IF{$||{\boldsymbol{a}}(k)||\leq\epsilon_{\mathrm{lo}}$ and $hit_{p}=\text{TRUE} $ and $hit_{v}=\text{FALSE} $} 
		\STATE$c_{\mathrm{lo}}=c_{\mathrm{lo}}+1$
		\STATE$hit_{p}=\text{FALSE} $, $hit_{v}=\text{TRUE} $
		\ENDIF
		\STATE $k=k+1$
		\ENDWHILE
		\STATE $\{Y_m\}=||{\boldsymbol{a}}(k_{\mathrm{start}}:k-1)||$
		\STATE$ \{t_m\}=t_{acc}(k-1)$
		\STATE $k_{\mathrm{start}}=k$
		\STATE $m=m+1$
		\STATE $c_{\mathrm{up}}=0$ and $c_{\mathrm{lo}}=0$
		\RETURN \{$ \boldsymbol{Y}$\}, \{$ \boldsymbol{t}$\}
	\end{algorithmic} 
	\label{GCDA}
\end{algorithm}
%
%
%
%

Peak and valley should be hit consequentially. However, we do not know which one comes first.  Hence, in order to make sure that both peaks and both valleys have been detected in order, two flags $hit_{p}$ or $hit_{v}$  are defined.  As soon as the first peak/valley has been hit, the corresponding  flag $hit_{p}$ or $hit_{v}$ is set to $\text{TRUE}$ and the other one will be $\text{FALSE}$. 
When both counters become larger than or equal to two, searching  will stop and  this part of the signal  will be considered as a gait cycle. Finally, all counters are reset to initial values and the same procedure is repeated for the rest of the signal. 
\begin{figure*}[!t] 
	\begin{center}
		\subfloat[ W1 \label{WFH}]{\includegraphics[width=0.5\textwidth]{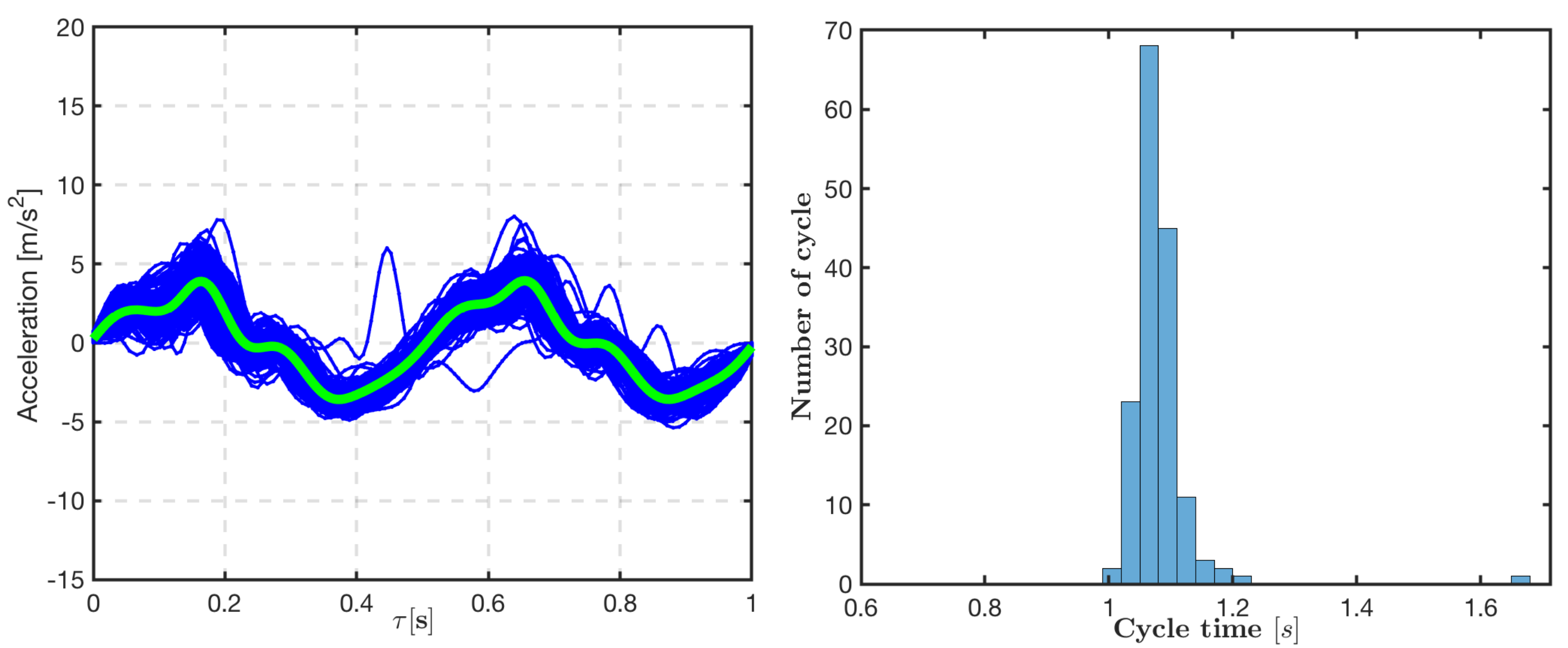}}\hfill
		\subfloat[ R1 \label{RFH}]{\includegraphics[width=0.5\textwidth]{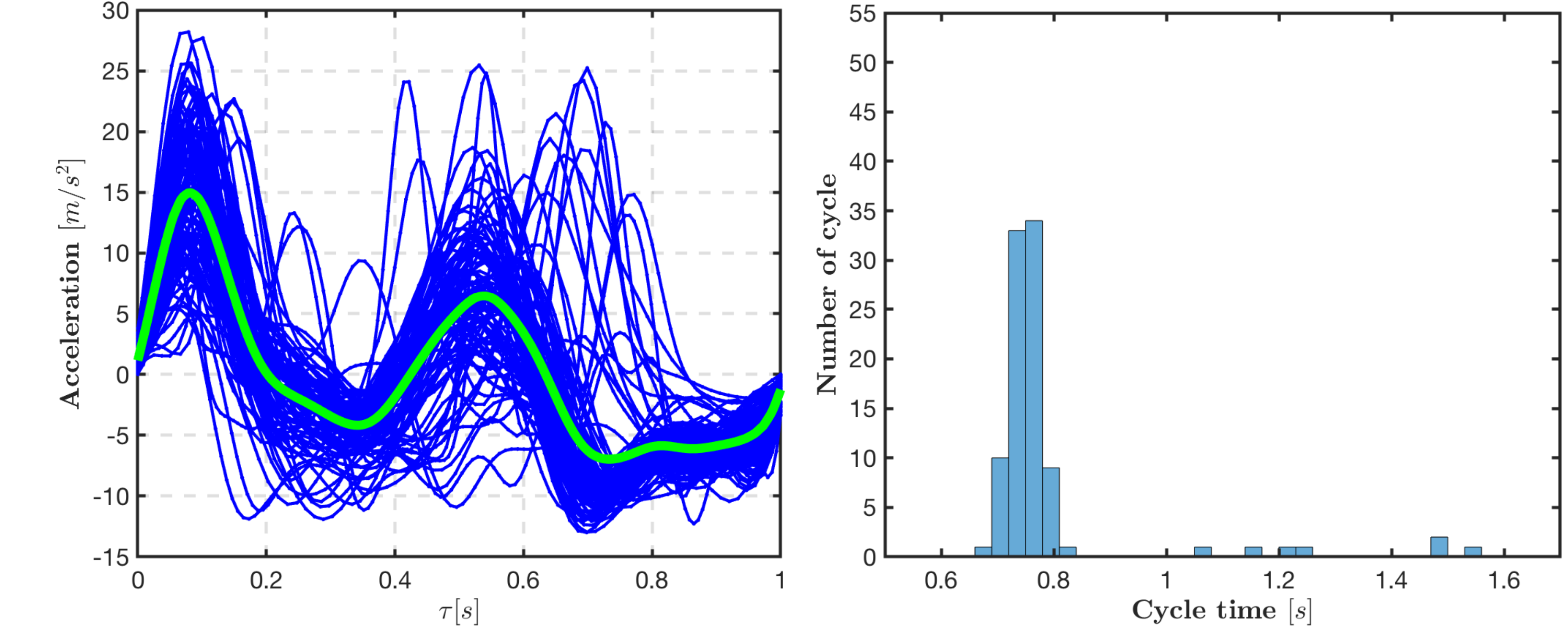}}\\
		
		\subfloat[ W2 \label{WSH}]{\includegraphics[width=0.5\textwidth]{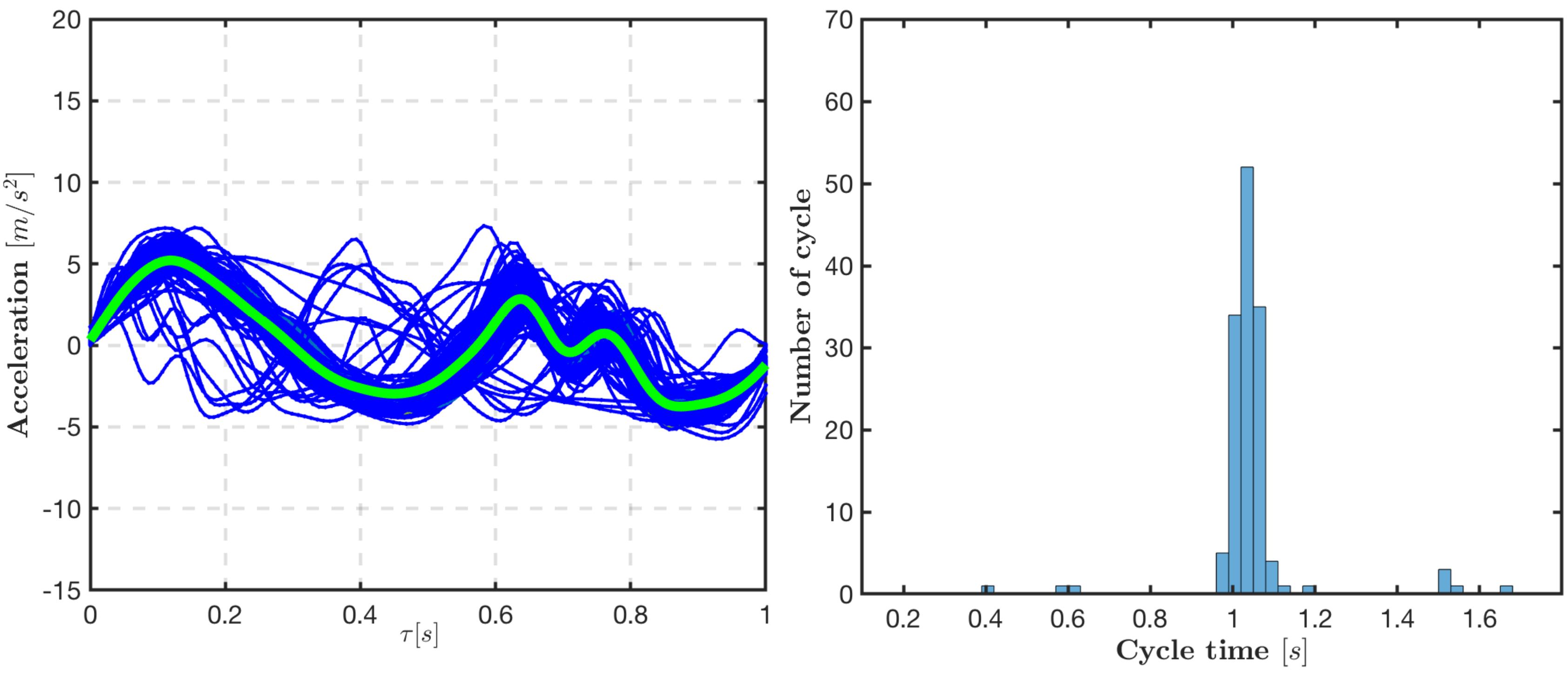}}\hfill
		\subfloat[ R2 \label{RSH}]{\includegraphics[width=0.5\textwidth]{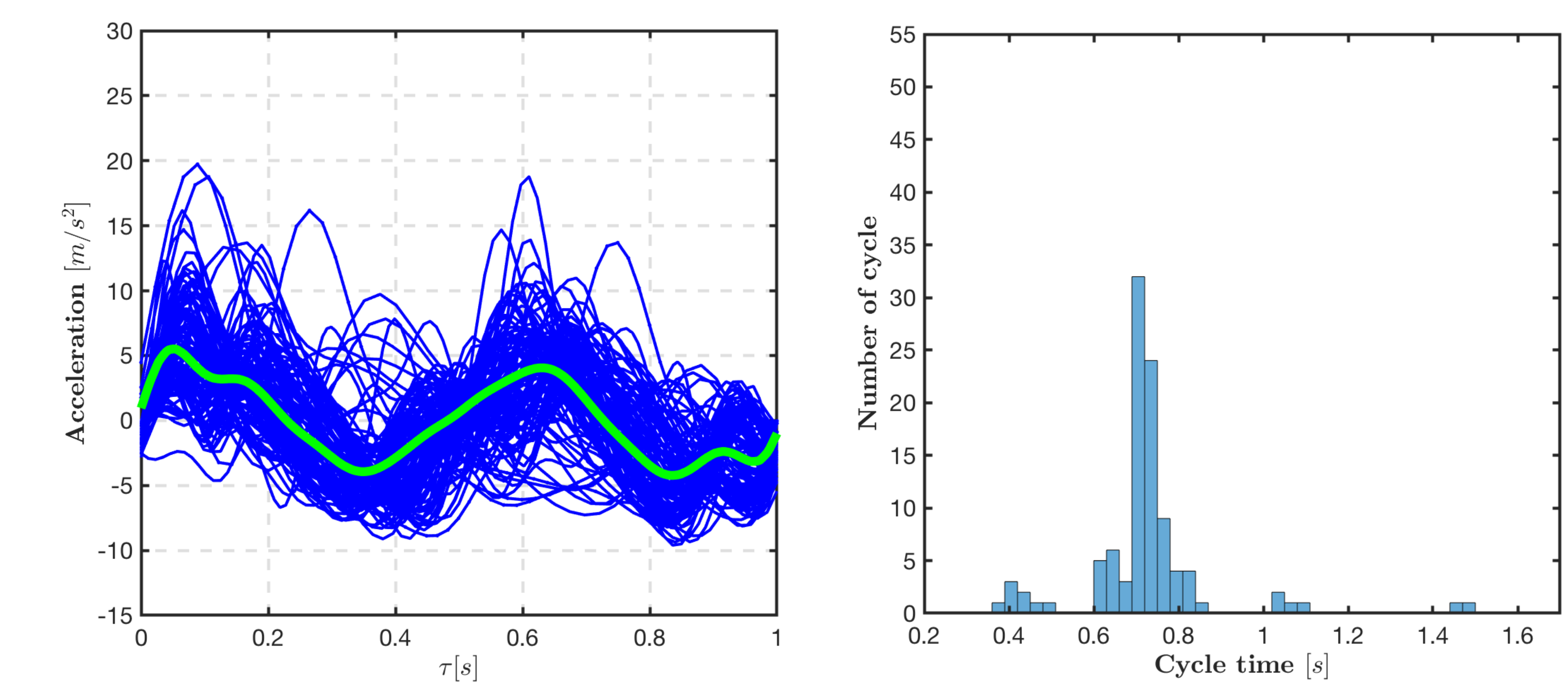}}\\
		\subfloat[ W3 \label{WFP}]{\includegraphics[width=0.5\textwidth]{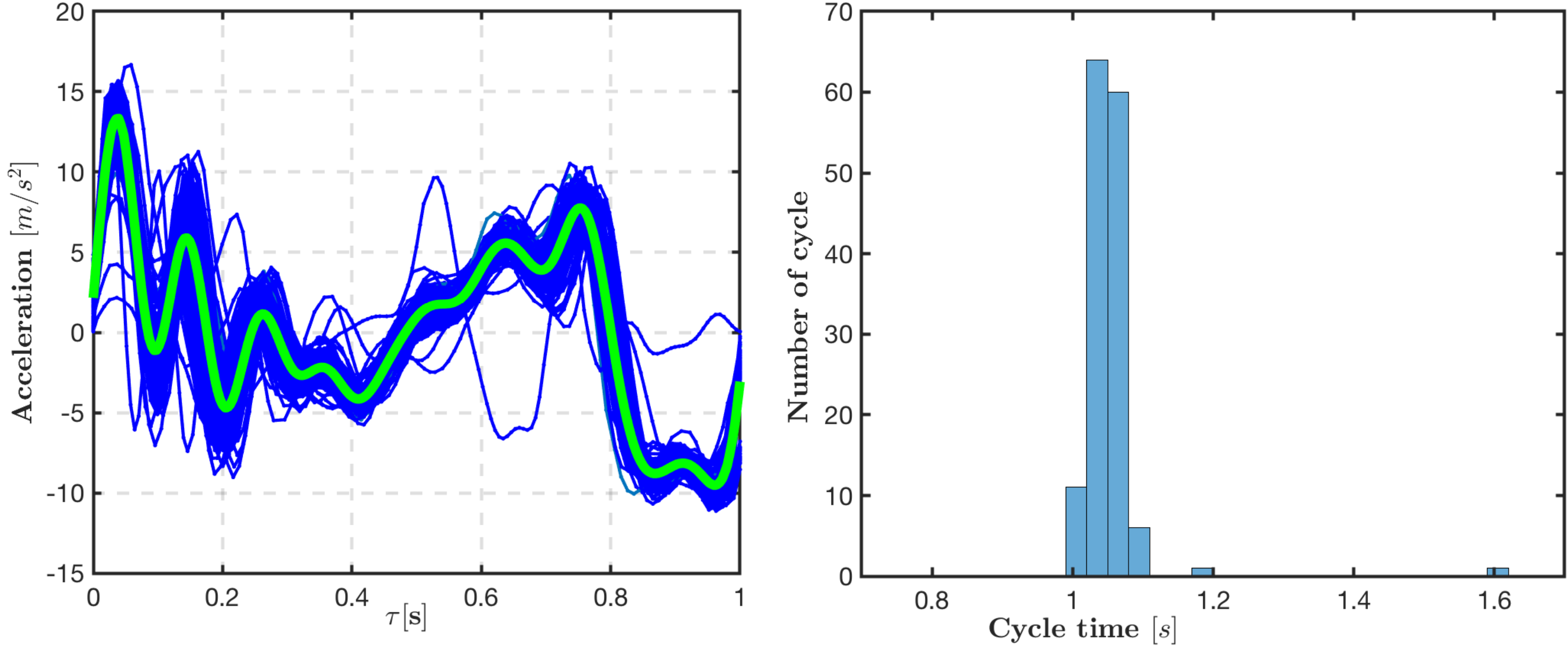}}\hfill
		\subfloat[ R3 \label{RFP}]{\includegraphics[width=0.5\textwidth]{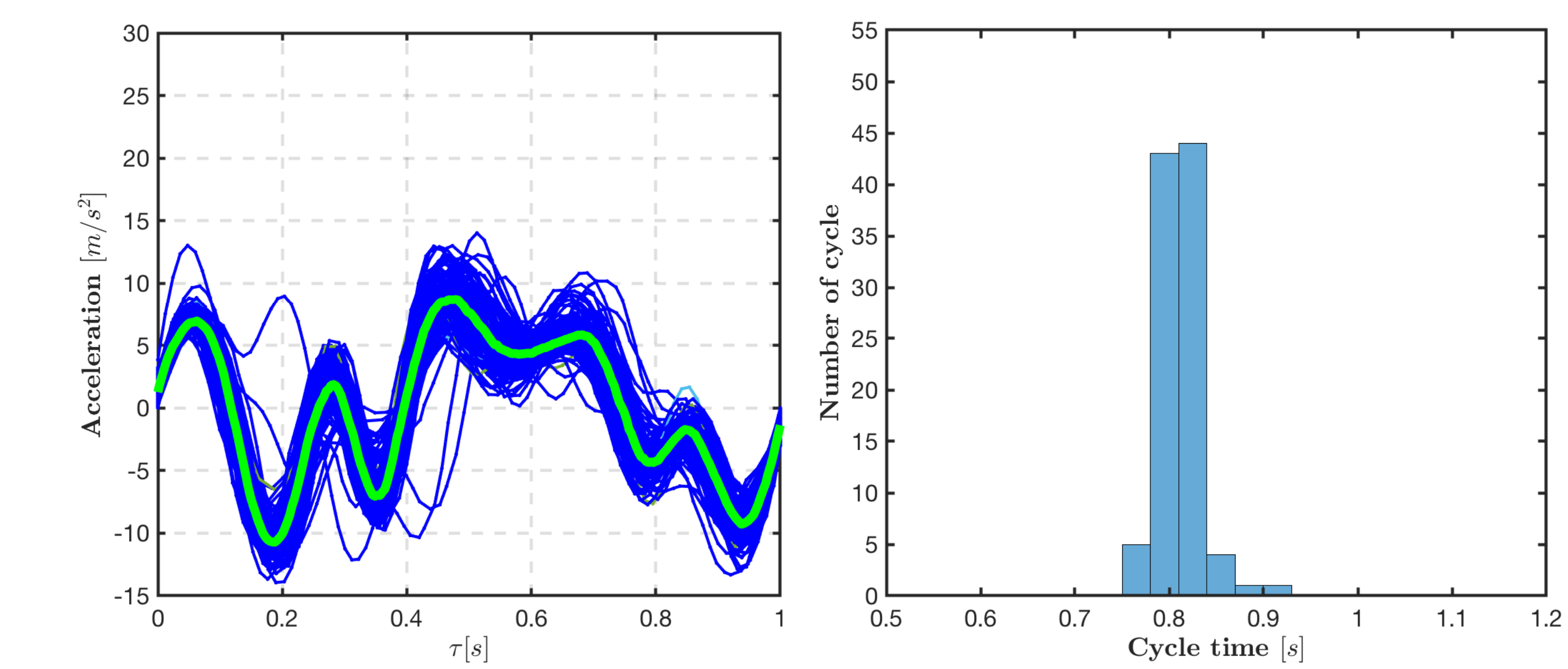}}\\
		\subfloat[ W4 \label{WBP}]{\includegraphics[width=0.5\textwidth]{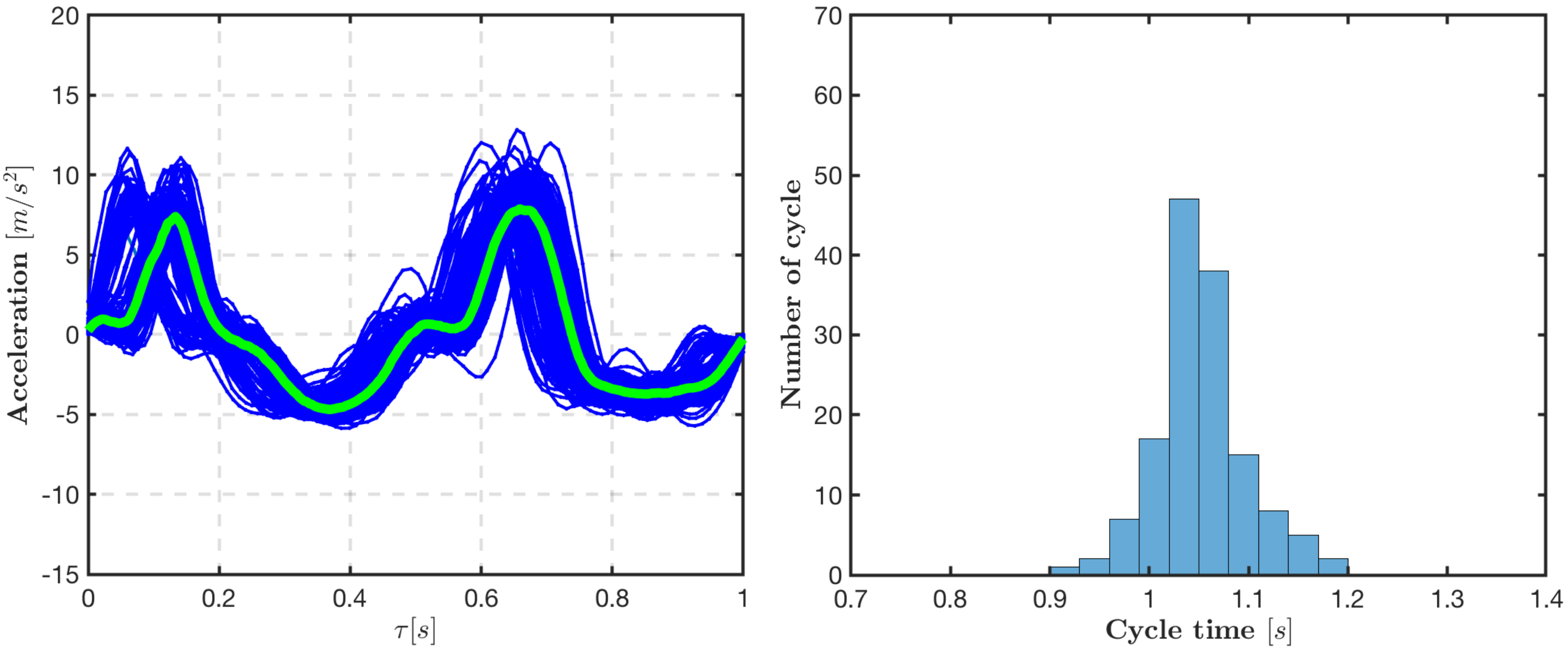}}\hfill 
		\subfloat[ R4 \label{RBP}]{\includegraphics[width=0.5\textwidth]{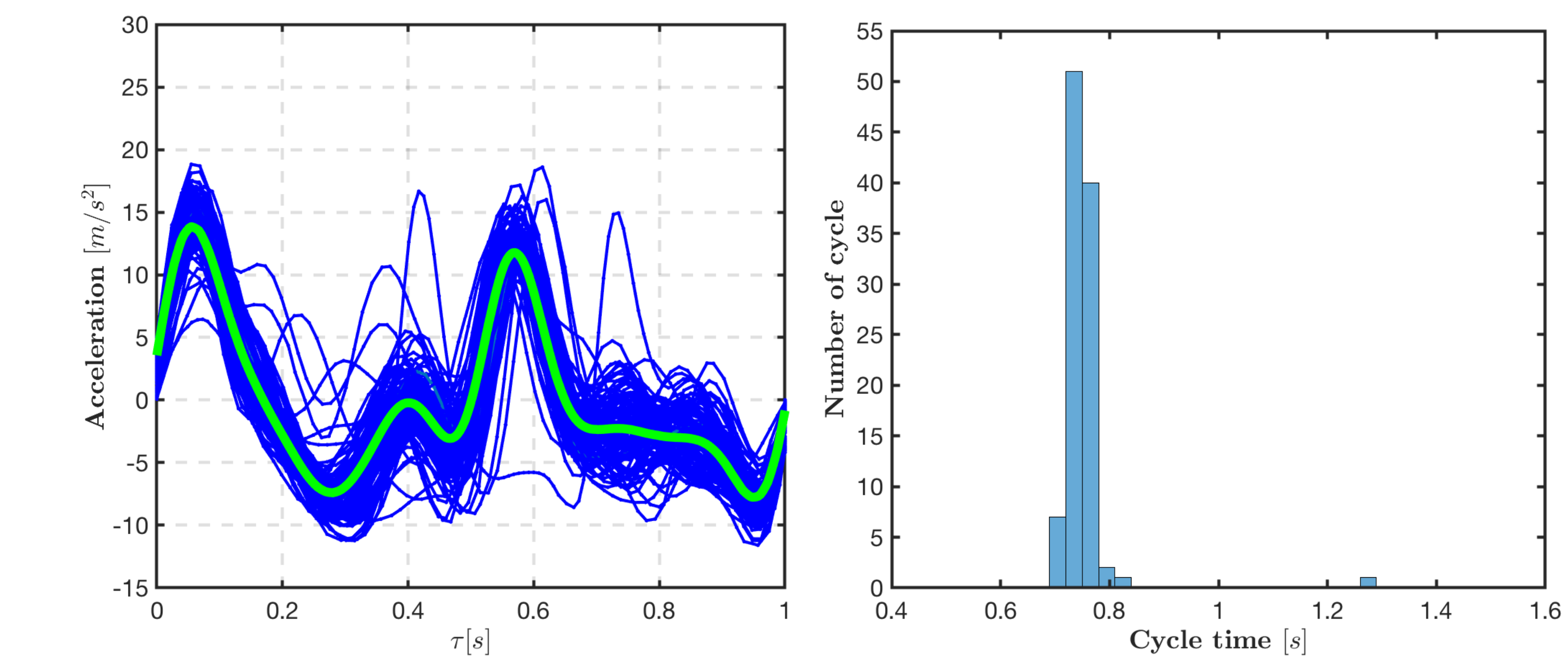}}
		
	\end{center}
	\caption{Results for all eight introduced scenarios in Table~\ref{tbl:classes}. In each sub-figure: On the right, detected gait cycles (blue lines) using Algorithm~\ref{GCDA} are presented and estimated reference signal $\boldsymbol{ \bar{g}}(\tau)$ is depicted on top of all detected gait cycles with a solid thick green line. On the left, the histogram for the gait cycle duration time is presented.}
	\label{RFmodesW}
\end{figure*}

Algorithm~\ref{GCDA} is applied to the experimental dataset (containing $8$ scenarios),  as  introduced in Sec.~\ref{sec:results} to get  the gait cycle segmentations. The hyperparameter values are set to $\epsilon_p=2$\,m/s$^2$ and $\epsilon_v=-2$\,m/s$^2$ for walking mode and $\epsilon_p=4$\,m/s$^2$ and $\epsilon_v=-5$\,m/s$^2$ for running mode. The extracted segments are illustrated in  Fig.~\ref{RFmodesW}. It is worth noting that all the gait cycle times are normalized such that $\tau \in [0,1)$.  

In this work, in order to extract the pattern of the gait cycle with better quality and to avoid bias drift, a pre-processing step is applied to the raw measured data. 
For this purpose, the signal is filtered through a fourth-order Butterworth filter with cut-off frequencies $[0.1, 10]$ Hz  to attenuate all frequencies outside the band-pass. 
The norm of the filtered signal is considered in this work in order to avoid unpredicted disturbances of the vertical acceleration that the orientation of the sensor may cause. 

 The initial gait signatures, $\bf{\bar g}({\tau})$, computed by \eqref{signature} are plotted over the detected gait cycles for all the scenarios and indicated with thick green line in Fig.~\ref{RFmodesW}. As the figures suggest, although a general pattern is visible for the extracted gait cycles in all scenarios, further tuning is required. For example, when the device is in swinging mode,  Fig.~\ref{WSH} and Fig.~\ref{RSH}, corresponding to walking  and running with swinging hand, the gait cycles are very noisy and there are some misdetected gait segments. Moreover, in ``W4'' scenario, presented in Fig.~\ref{WBP}, corresponding to walking with backpack device mode, there is  some shifting. In the running activity mode the patterns are quite noisy and the lengths of the cycle times also vary as shown in the histograms corresponding to running gait modes.

Unexpected behaviors in the motion and the device modes are unpredictable and always exist in pedestrians’ daily activities. Hence, these classical algorithms such as the one given by Algorithm~\ref{GCDA} should be complemented by more advanced methods. 

The introduced optimal segmentation in Sec.~\ref{sec:optimization}, suggests a solution to take care of unexpected behaviors and reduce the rate of error. 
The performance of the proposed method is evaluated on a large dataset and presented in Sec.~\ref{sec:results}. 

\section{Data Reduction with Fourier Series}
\label{sec:FS}
In this section,  we are looking for a low order approximation, which can be recasted as a least squares (LS) estimation problem using a linear regression framework. 

Given our sequence of step times and the measurement noise we have for each $\hat{g}_m(\tau_l)$, we strive to find a parametric model, using a linear regression framework, of the average $\bar{g}(\tau_l)$. We show that the parameters of the fitted model provide a useful set of features for future classification purposes.
%

In order to extract a low dimensional feature vector for the gait cycle using estimated signatures,  we apply FS expansions to the gait segment from the sampled version of the gait cycle $\boldsymbol{\hat{G}}_m[l]$, $l=[0,1,\dots,L-1]$ 
\begin{align}
\boldsymbol{\hat{G}}[l]
= \sum_{k=0}^{K-1} a_k \cos(2\pi\tau_l k) + b_k \sin(2\pi\tau_l k),
\label{FS}
\end{align}
where the parameter set $\{a_k,b_k\}_{k=1}^K$ forms the feature space used to identify each particular gait mode and $K$ is the model order. As in any other regression model, the trade-off between model complexity and accuracy cannot be neglected. 
In order to estimate the FS coefficients, for each model order $K$, the Fourier series expansion~\eqref{FS} is considered as a linear model given by
\begin{align}
\boldsymbol{\bar{g}}= \boldsymbol{H}_K\boldsymbol{ \theta}_K,
\label{FSlinear}
\end{align}
where $\boldsymbol{{\theta}}$ is a vector containing all unknown coefficients, $\boldsymbol{{\theta}}_K^{2K\times 1}=[a_0, a_1,\dots,a_{K-1},b_0,b_1,\dots,b_{K-1}]^\top$, and $\boldsymbol{\bar{g}}^{L\times 1}=[\boldsymbol{\bar{g}}(\tau_0),\boldsymbol{\bar{g}}(\tau_1),\dots,\boldsymbol{\bar{g}}(\tau_{L-1})]^\top$. $\boldsymbol{H}_K^\top \in \mathbb{R}^{2K\times L}$ can then be given by 
\begin{align*}
\boldsymbol{H}_{K}^{\top}{=}{\begin{bmatrix}
1 			& \cos(2\pi\tau_1)				&\dots  & \cos(2\pi\tau_{L-1})			\\ 
\vdots & \vdots						  	 &\dots  & \vdots						 		  \\ 
1			&  \cos(2\pi\tau_1(K-1))	&\dots  &  \cos(2\pi\tau_{L-1}(K-1))  \\ 
0			& \sin(2\pi\tau_1) 				& \dots & \sin(2\pi\tau_{L-1})				\\ 
\vdots	& \vdots 							&\dots  & \vdots 									\\ 
0			& \sin(2\pi\tau_1(K-1))		&\dots  & \sin(2\pi\tau_{L-1}(K-1)) 	  \\ 
\end{bmatrix}}{.}
\end{align*}
The solution to the problem is obtained by finding the following optimization problem
\begin{equation}
\begin{aligned}
\boldsymbol{\hat \theta}=
& \underset{\boldsymbol{ \theta}}{\text{minimize}}
& & V^{LS}(\boldsymbol{ \theta}),
\end{aligned}
\label{LSopt}
\end{equation}
where 
\begin{equation}
V^{LS}(\boldsymbol{ \theta}_K)=(\boldsymbol{\bar{g}}-\boldsymbol{H}_K\boldsymbol{ \theta}_K)^\top(\boldsymbol{\bar{g}}-\boldsymbol{H}_K\boldsymbol{ \theta}_K).
\end{equation}
\begin{figure}[!t] 
	\centering
	\includegraphics[width=0.8\columnwidth]{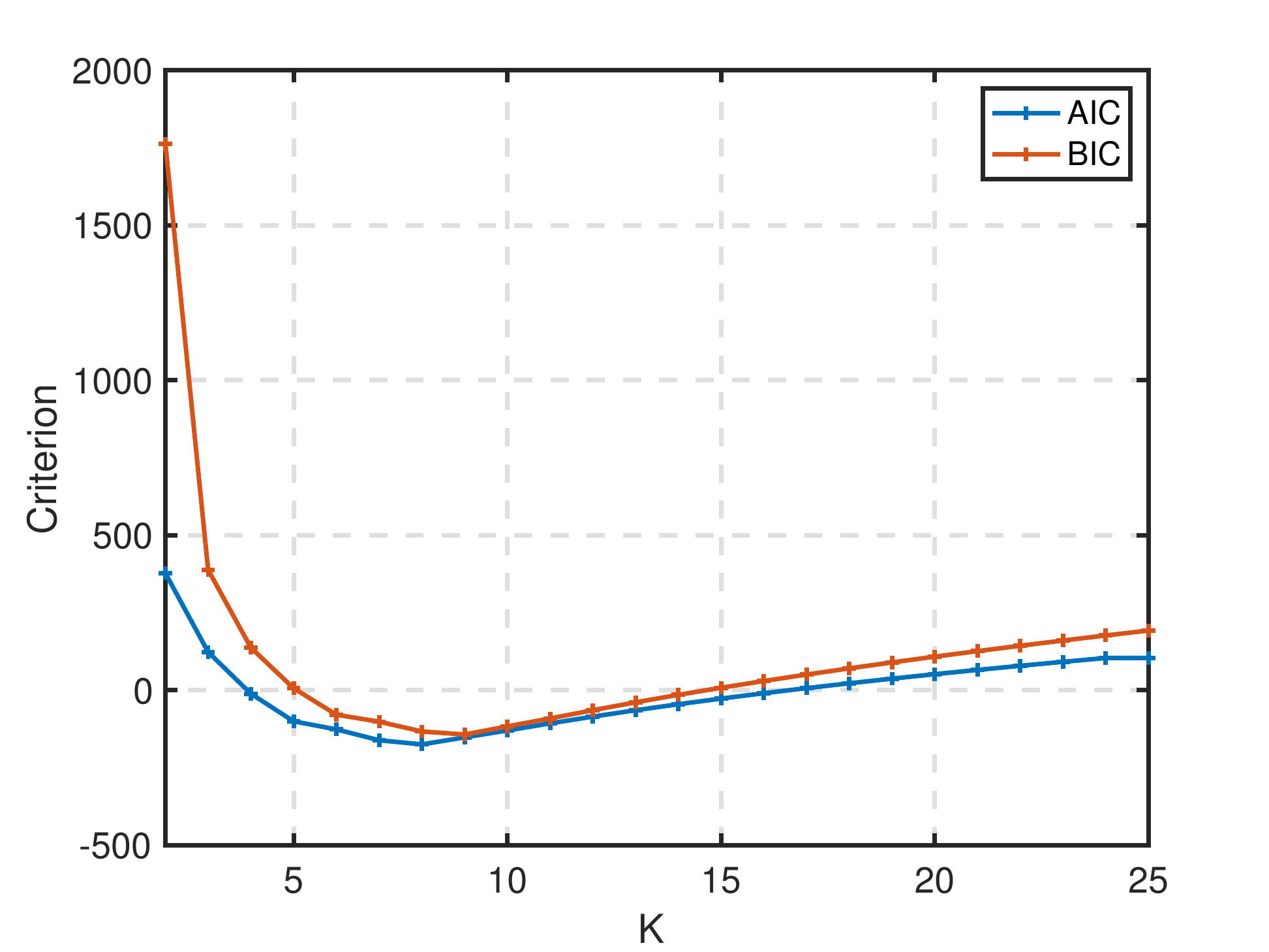} 
	\caption{Model order selection.}
	\label{fig:AIC}
\end{figure}
Finally, the closed form solution $\boldsymbol{\hat \theta}$ is given by 
\begin{align}
\boldsymbol{\hat\theta}_K^{}=(\boldsymbol{H}_K^\top \boldsymbol{H}_K^{})^{-1}\boldsymbol{H}_K^\top\boldsymbol{\bar g}.
\end{align}
\begin{figure}[!t] 
	\begin{center}
		\subfloat[FS approximate of the signature generated by~\eqref{FS} for K=8.\label{FSplot}] {\includegraphics[width=0.49\columnwidth]{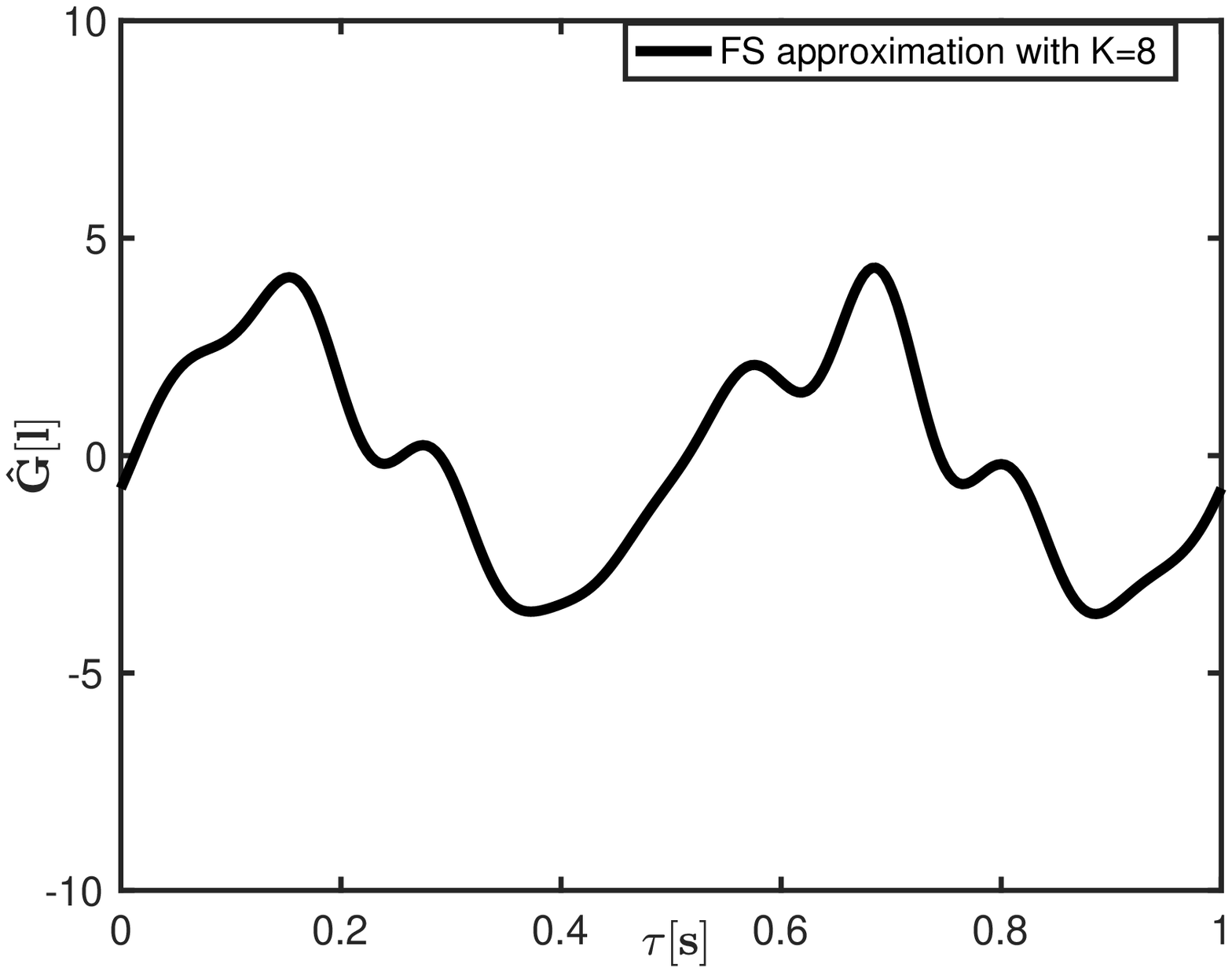}} \hfill 
		\subfloat[Estimated signature using~\eqref{signature}.\label{Sig}]
		{\includegraphics[width=0.49\columnwidth]{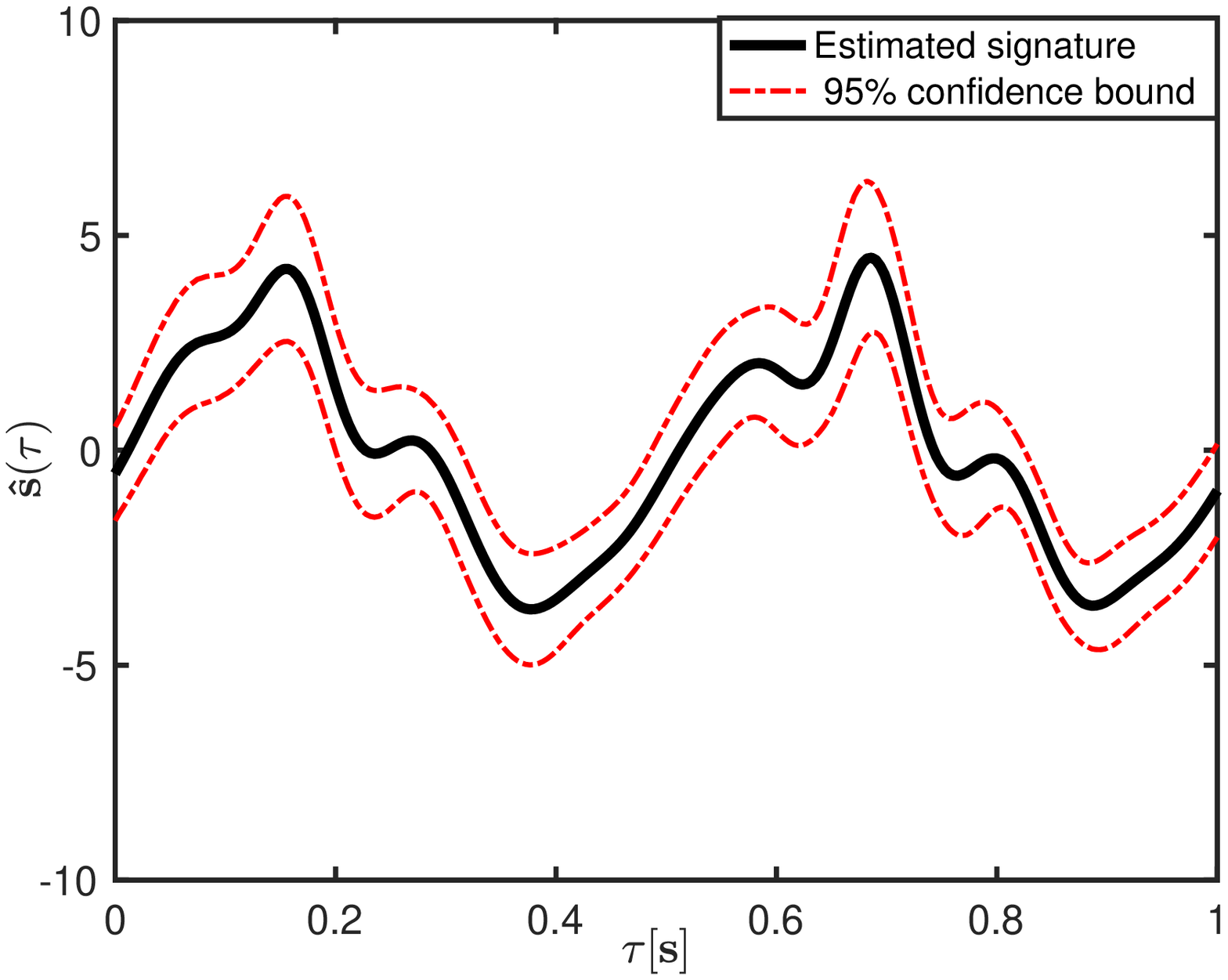}}
	\end{center}
	\caption{Comparison between the estimated signature and its FS expansion.}
	\label{confbound}
\end{figure}

To find the best order model in FS expansion the gait cycle is modeled for $K\in[1,\ldots,25]$. Then, the FS coefficients, for each $K$ are estimated and further evaluated by two well-known model selection criteria: Akaike information criterion (AIC) and Bayesian information criterion (BIC). Both AIC and BIC add a penalty term to their objective function. The difference between the two criteria is that BIC imposes a greater penalty for the number of parameters compared to AIC~\cite{Book:Fabozzi2014}. For a sample of size $L$, for each model order $K$, AIC and BIC are defined as

\begin{subequations}
	\begin{align}
	AIC &= -2\log p(\boldsymbol{\bar g} \mid \boldsymbol{\hat \theta}_K) +2K,\\
	BIC &=  -2\log p(\boldsymbol{\bar g} \mid \boldsymbol{\hat \theta}_K) +K\log L,
	\end{align}
\end{subequations}
where $\mathrm{p(\boldsymbol{\bar g} \mid \boldsymbol{\hat \theta}_K)} =\mathcal{N}(\mathrm{\boldsymbol{H}_K\boldsymbol{\hat \theta_K}}, \mathrm{\boldsymbol{H}_K
	(\boldsymbol{H}_K^\top \boldsymbol{H}_K^{})^{-1}_K\boldsymbol{H}_K^\top})$. 

As Fig.~\ref{fig:AIC} suggests, $K=8$ and $K=9$ are  the suitable model orders for the gait segments presented in Fig.~\ref{manualGC} according to AIC and BIC,  respectively. While either of the two orders could be selected as the suitable model order for this scenario, we select the lower order, $K=8$. 
Hence, the corresponding signatures can be  generated by only incorporating  $2K=16$ estimated coefficients of the FS, $\boldsymbol{\hat \theta}$,  into~\eqref{FS},  as shown in Fig.~\ref{FSplot}. Fig.~\ref{Sig} presents  the gait signature $\boldsymbol{\bar g}(\tau)$, and its  95\% confidence bound. As this figure shows, the averaging error has quite wide variance. 

\begin{figure}[!t] 
\centering
\includegraphics[width=0.8\columnwidth]{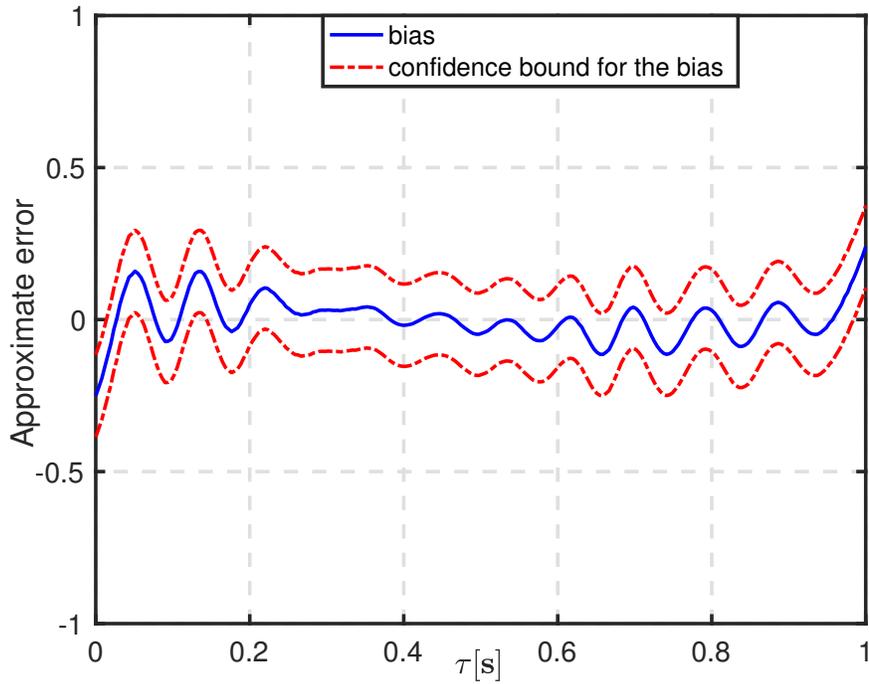}
\caption{The approximation error and the $95\%$ confidence bound.}
\label{Bias}
\end{figure}

To evaluate the encoded signal by the estimated FS coefficients with $16$ components, the approximate error between $\boldsymbol{\hat  G}[l]$ and  $\boldsymbol{\bar g}(\tau)$ are computed and illustrated in  Fig.~\ref{Bias}. As the result indicates, the approximation error is quite small with narrow confidence bounds. That is, the approximation error introduced by the low order FS expansion is negligible compared to the averaging error.
Subsequently, this low-dimensional feature vector together with the final least square cost value and the step time variations could provide a useful set of features for future classification purposes.

%
%
%
\section{Experimental Results}
\label{sec:results}
In this section, the proposed method introduced in Sec.~\ref{sec:Opt} is evaluated on several experimental data. Additionally, the data collection setup, together with devices and all considered scenarios, is described in detail. 

\subsection{Data Description}
\label{sec:DD}
In order to evaluate the performance of the proposed method, an extensive measurement campaign with different human motion modes and device poses has been designed. 
\begin{table}[!b]
	\centering
	\caption{Experimental scenarios.}
	\label{tbl:classes}
	\begin{tabular}{c|c|c}

		\backslashbox{\textbf{\begin{tabular}[c]{@{}c@{}}Device mode\end{tabular}} }{\textbf{\begin{tabular}[c]{@{}c@{}}Motion  mode\end{tabular}} }            & \textbf{\begin{tabular}[c]{@{}c@{}}Walking\\ (W)\end{tabular}} 
		& \textbf{\begin{tabular}[c]{@{}c@{}}Running\\ (R)\end{tabular}} \\ \hline
		\textbf{Fixed hand (1)}    		  	&  W1              &  R1          \\ 
		\textbf{Swinging hand (2)} 		 &  W2             &  R2          \\ 
		\textbf{Pocket (3)}       			  &  W3             &  R3        	\\ 
		\textbf{Backpack (4)}      			&  W4             &  R4            \\ 
	\end{tabular}
\end{table}
The sensor fusion Android app~\cite{sensorfusionapp,hendebyGW:2014} installed on a Nexus 5 was used to log accelerometer and gyroscope measurements with a sampling rate of $100~$\,Hz. All the measurements were collected over the same trajectory, which was $249$\,m  in length with four sharp corners in a parking lot at Link\"oping University. 
Several subjects with different attributes (gender, height and weight) participated in the experiment.
The data was collected for multiple human activities and device modes. Table~\ref{tbl:classes} summarizes all the experimental scenarios. To simplify referring to each of these scenarios, Table~\ref{tbl:classes} also assigns a specific symbol to each of them~\cite{Conf:kasebzadeh2017}. 
\subsection{Performance Evaluation}
%

As the first step, the gait cycles for all the scenarios are detected using Algorithm~\ref{GCDA} and illustrated in Fig.~\ref{RFmodesW} together with estimated signatures which are computed based on~\eqref{signature}. 
%

%
%
\begin{figure}[!t] 
	\begin{center}
		\includegraphics[width=0.8\columnwidth]{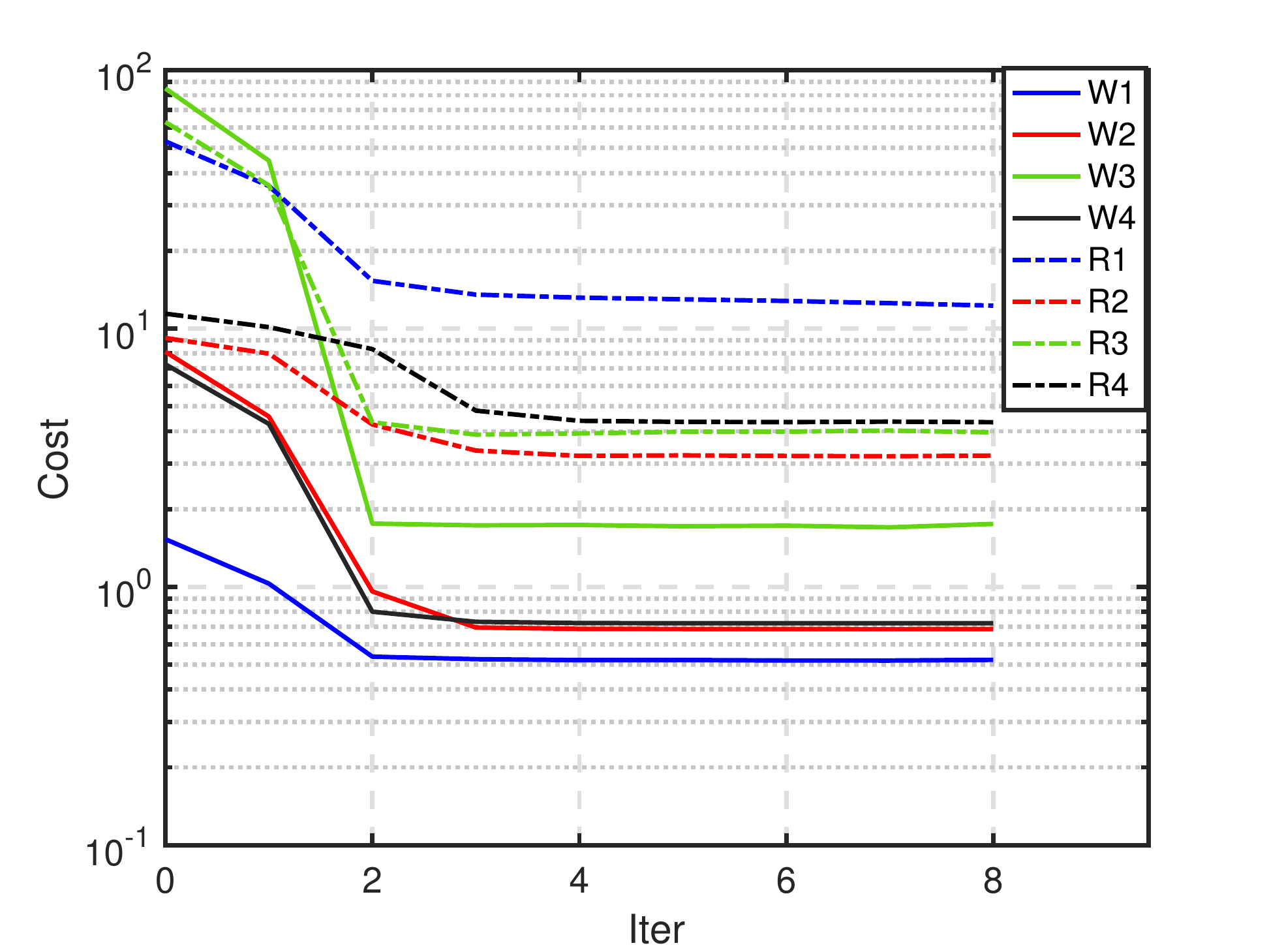}
	\end{center}
	\caption{Evaluation $V(t_0{:}t_M)$ in~\eqref{eq:genopt3} for eight iterations considering all scenarios in Table~\ref{tbl:classes}.}
	\label{costFun}
\end{figure}
\begin{figure}[!t] 
		\begin{center}
			\subfloat[ Walking \label{Wbox}]{\includegraphics[width=0.5\textwidth]{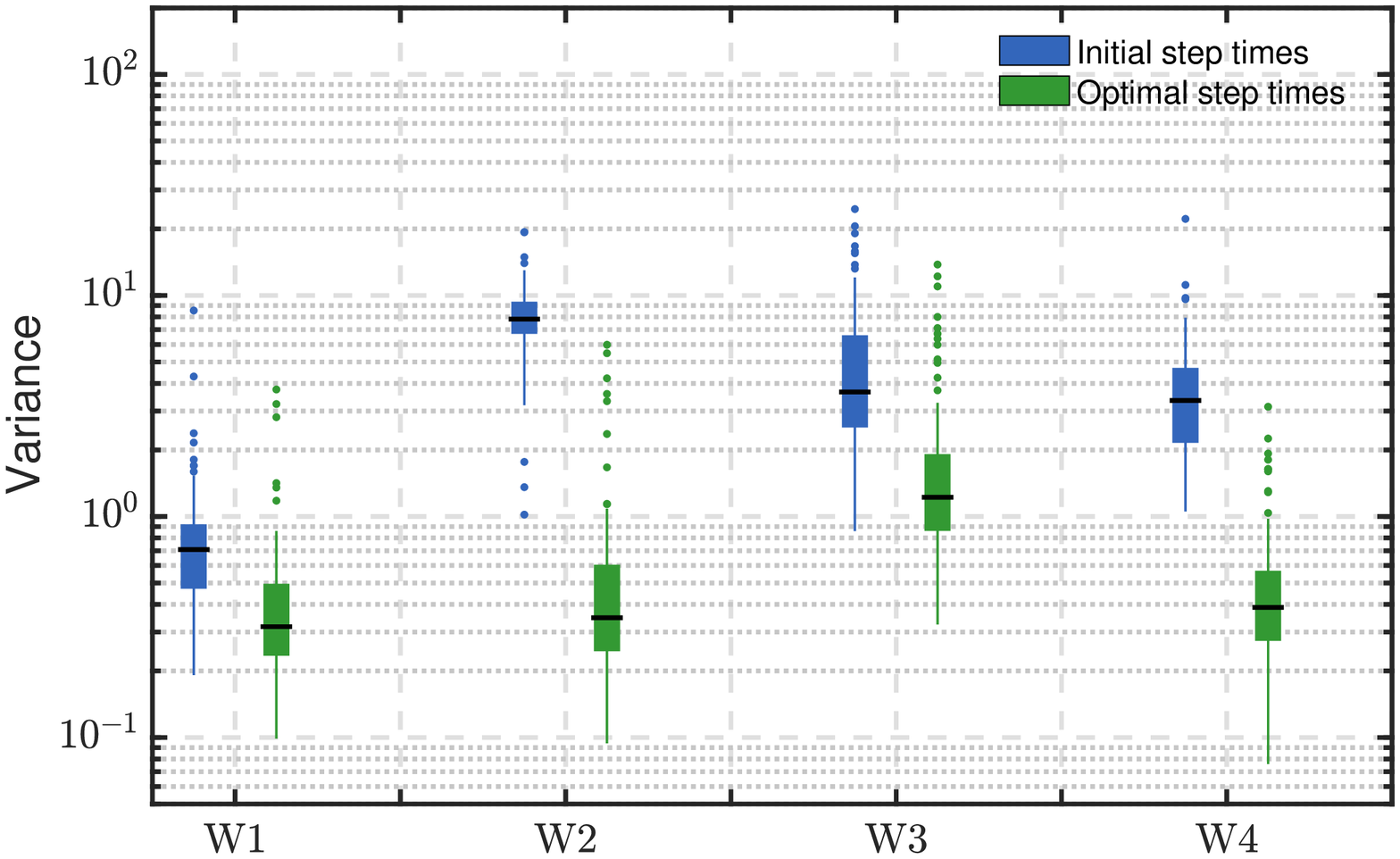}}\hfill
			\subfloat[ Running \label{Rbox}]{\includegraphics[width=0.5\textwidth]{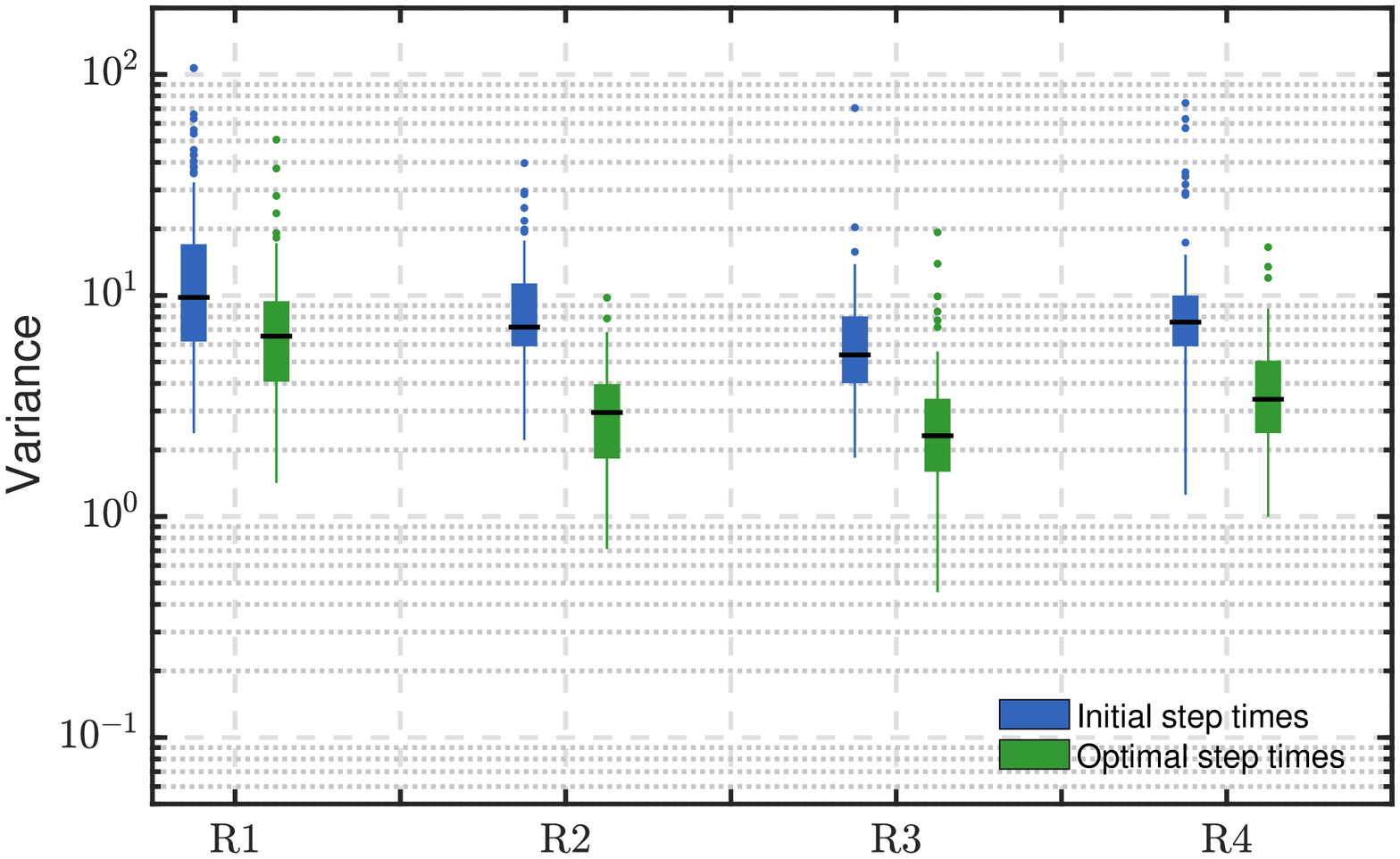}}\\
			
		\end{center}
		\caption{Variance of initial (blue) and tuned (green) gait cycles for all scenarios in Table~\ref{tbl:classes}.}
		\label{boxplot}
\end{figure}
Then, 
in order to fine-tune the gait segments extracted using the classical threshold-based algorithm, the proposed Algorithm~\ref{Opt_Alg} introduced in Sec.~\ref{sec:optimization} is employed.  The minimum and maximum of the most probable gait cycle times for each scenario, based on  the histograms  in Fig.~\ref{RFmodesW}, are used as a reference to define the upper and lower bounds in the optimization problem. The bounds are set to $\epsilon_{lo}=0.5$\,s and $\epsilon_{up}=1.4$\,s.

Initial gait cycle times, obtained from Algorithm~\ref{GCDA}, are  used to initialize the gait segmentation algorithm. Moreover, the initial signatures $\bf{\bar g}({\tau})$ to be used in Algorithm~\ref{Opt_Alg} are estimated by~\eqref{signature} using the initial gait cycle times. It is worth noting that after each iteration, the signature $\bf{\bar g}({\tau})$ will be updated by the tuned  gait cycles. 
Fig.~\ref{costFun} presents the evaluated cost function for eight iterations of the optimization progress considering all scenarios introduced in Table~\ref{tbl:classes}.  As the results indicate, in all scenarios after four iterations the cost function converges to the optimal values. 

In order to evaluate how much the gait cycles are improved, the variance of all initial gait cycles with initial signature are compared with all optimal gait cycles with updated signatures. The distribution of the obtained variances is shown in Fig.~\ref{boxplot}. The box levels are $5 \%$, $25 \%$, $50 \%$,
$75 \%$, and $95\%$ quantiles and the asterisks show outlier values. As the figure suggests, for all scenarios the optimal variances are improved significantly and the mean of the variance of the optimal gait cycles is decreased notably. By comparing the walking mode in Fig.~\ref{Wbox} the running mode one in Fig.~\ref{Rbox}, it can be verified that running involves more unexpected movement for the device, hence there are more disturbances and it has a higher variance in total. 

%
\begin{figure}[!t] 
	\begin{center}
		\subfloat[ W2 tuned gait cycles\label{W2final}]{\includegraphics[width=0.45\textwidth]{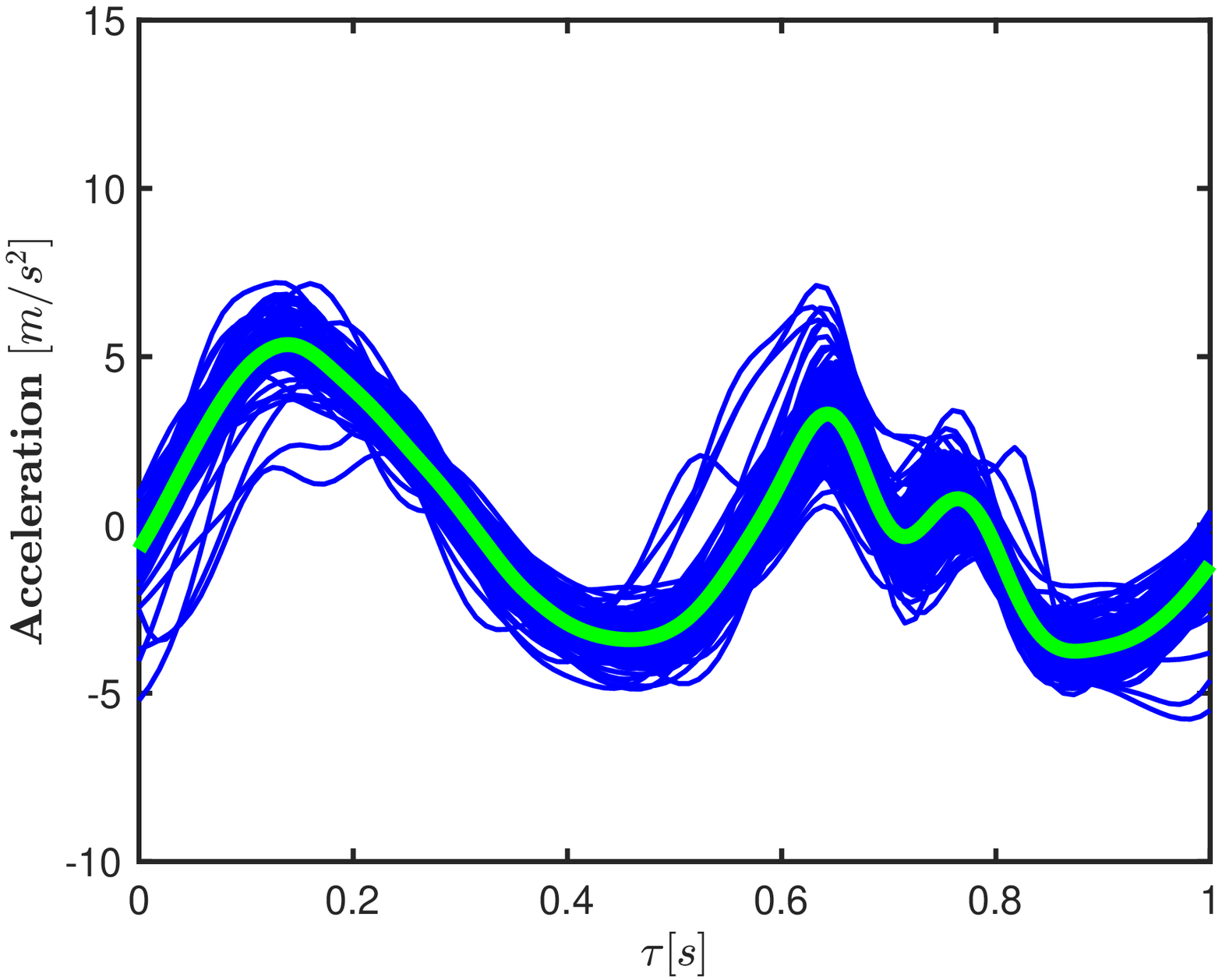}}
		\subfloat[ W2 step time duration \label{w2histfinal}]{\includegraphics[width=0.45\textwidth]{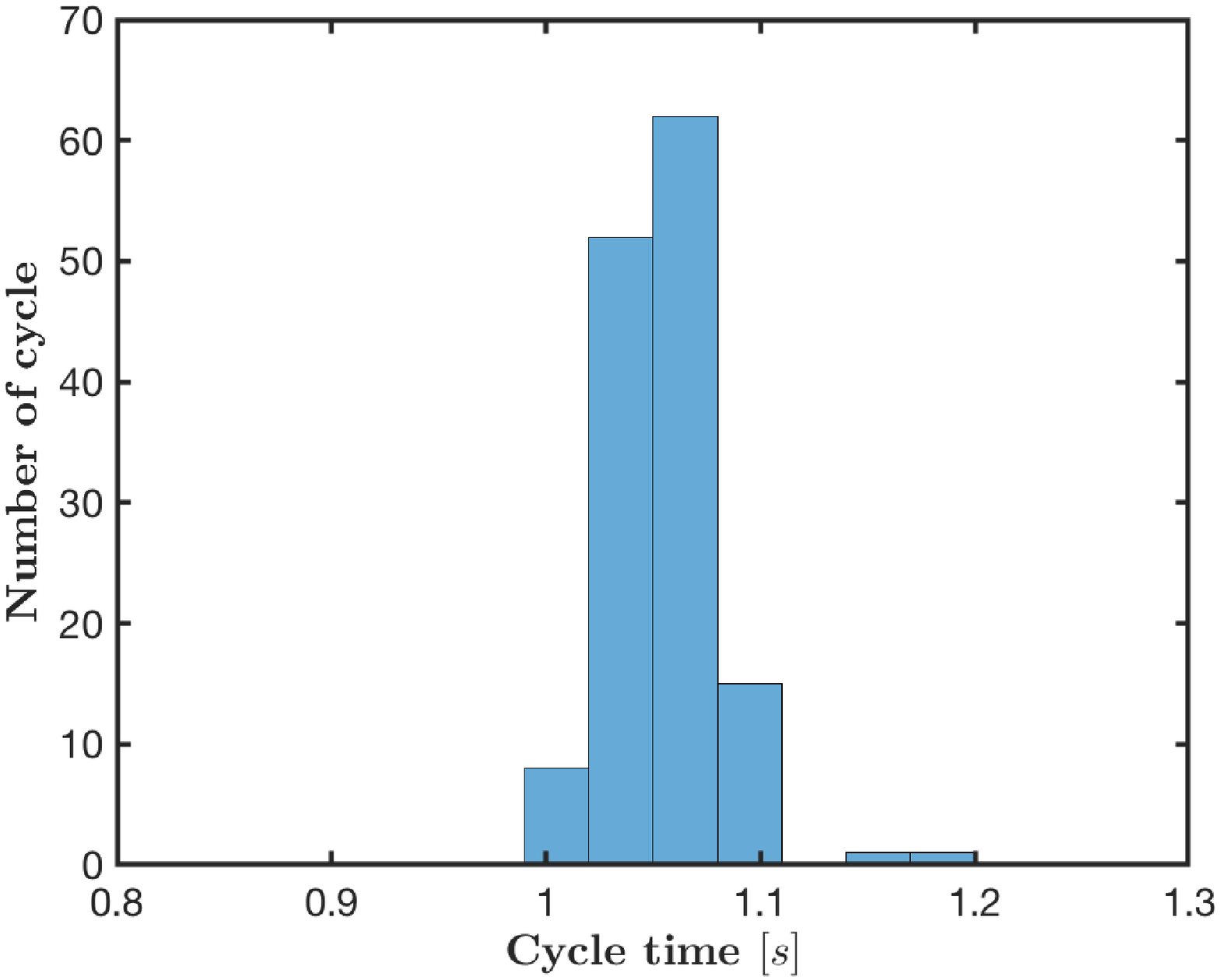}}\\
			\subfloat[ R2 gait cycles \label{R2final}]{\includegraphics[width=0.45\textwidth]{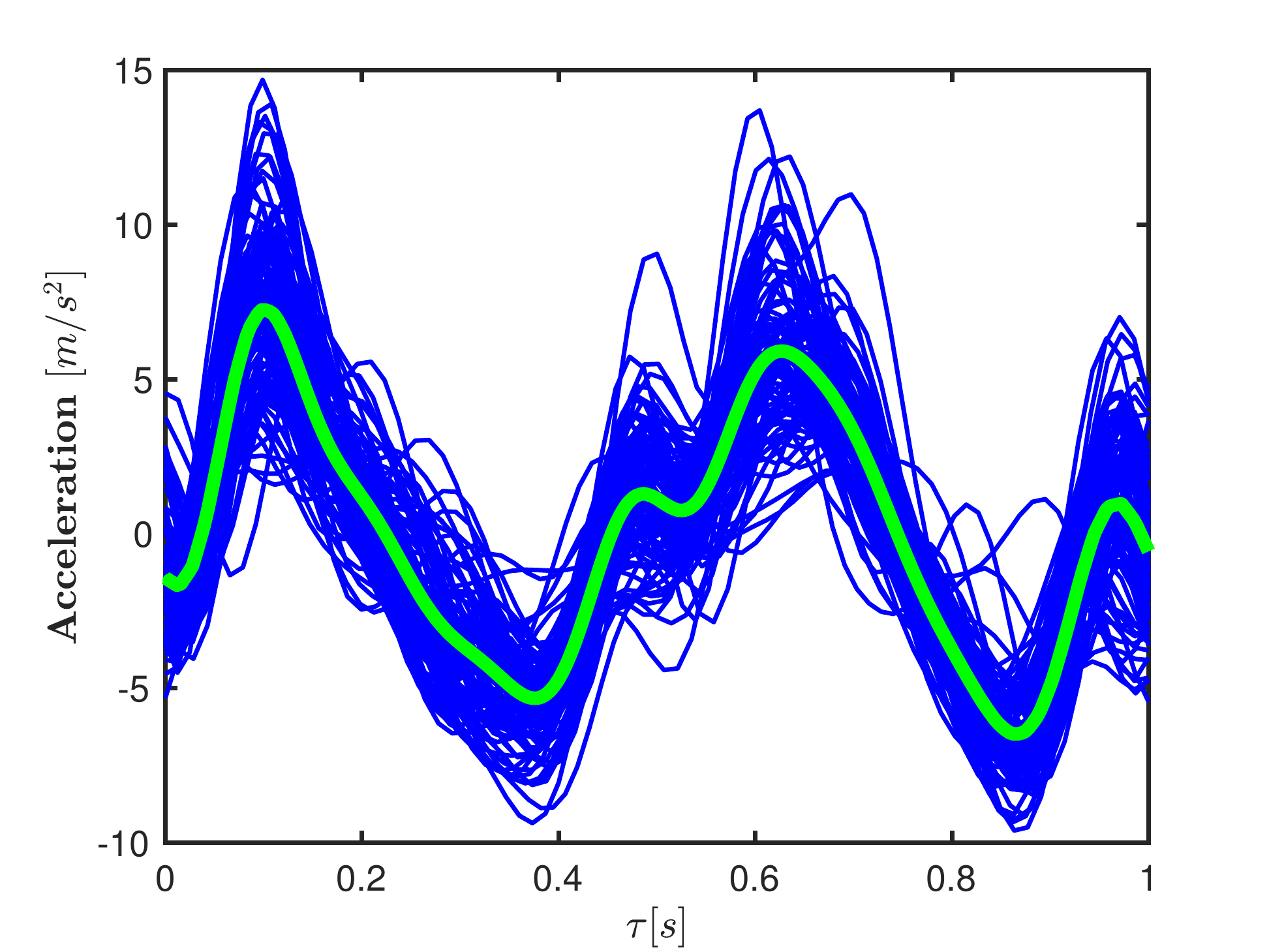}}
		\subfloat[ R2 step time duration \label{R2histfinal}]{\includegraphics[width=0.45\textwidth]{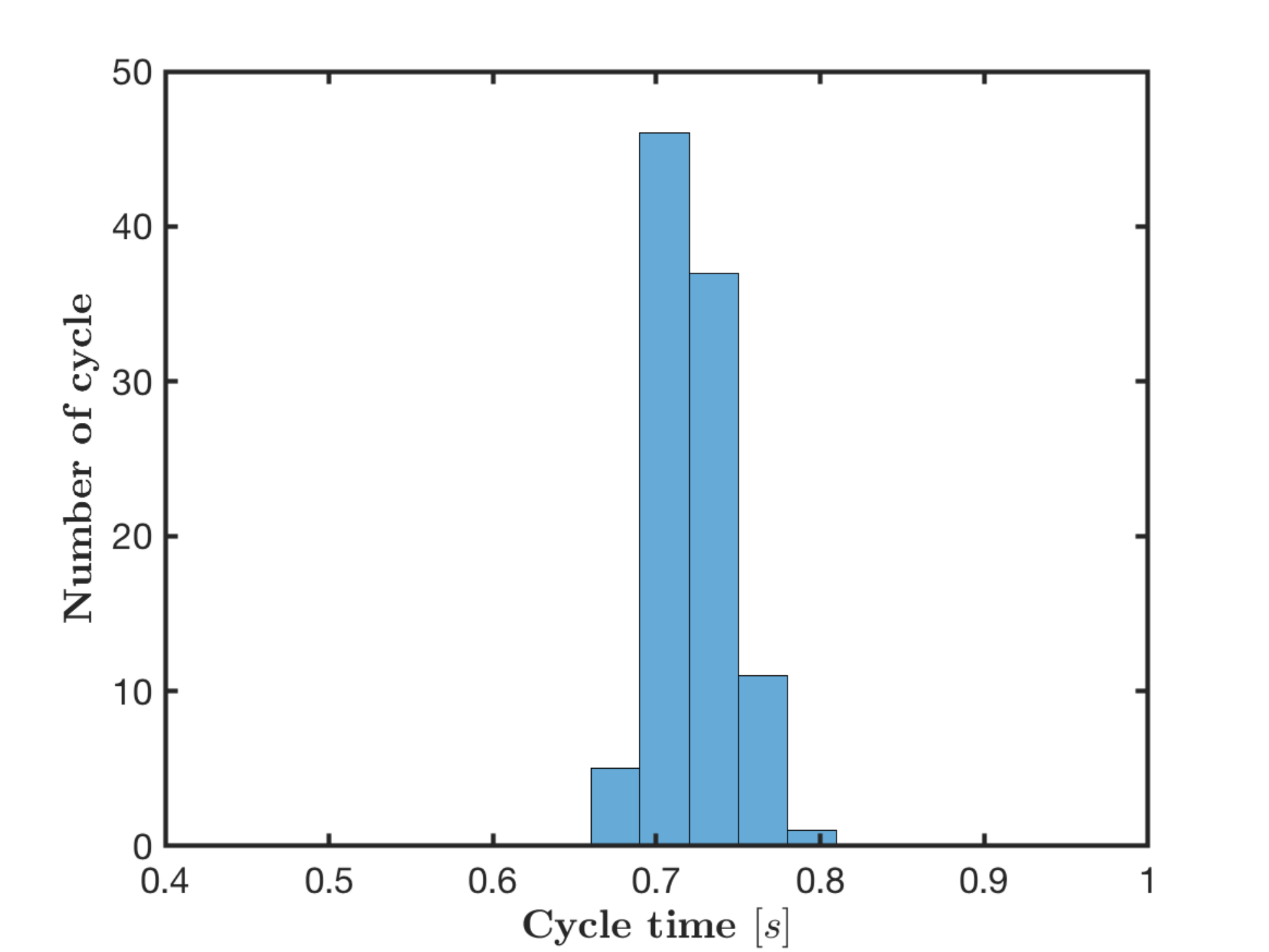}}\\
	\end{center}
	\caption{Optimized gait cycles with updated signatures  for ``W2'' and ``R2''  introduced scenarios in Table~\ref{tbl:classes}.}
	\label{optimizedWR2}
\end{figure}

In order to give a better illustration of the performance of the proposed algorithm, we further examine ``W2'' and ``R2'' scenarios. The tuned gait cycles for these two scenarios are presented in Fig.~\ref{optimizedWR2}. Comparing the results with initial gait cycles, see Fig.~\ref{WSH} and~\ref{RSH}, we realize that all the misdetected gait cycles are properly detected, as indicated in the histograms Fig.~\ref{w2histfinal} and~\ref{R2histfinal},  and all the gait cycles are perfectly tuned.

\begin{figure}[!t] 
	\begin{center}
		\subfloat[ W1 \label{WFHshat}]{\includegraphics[width=0.35\textwidth]{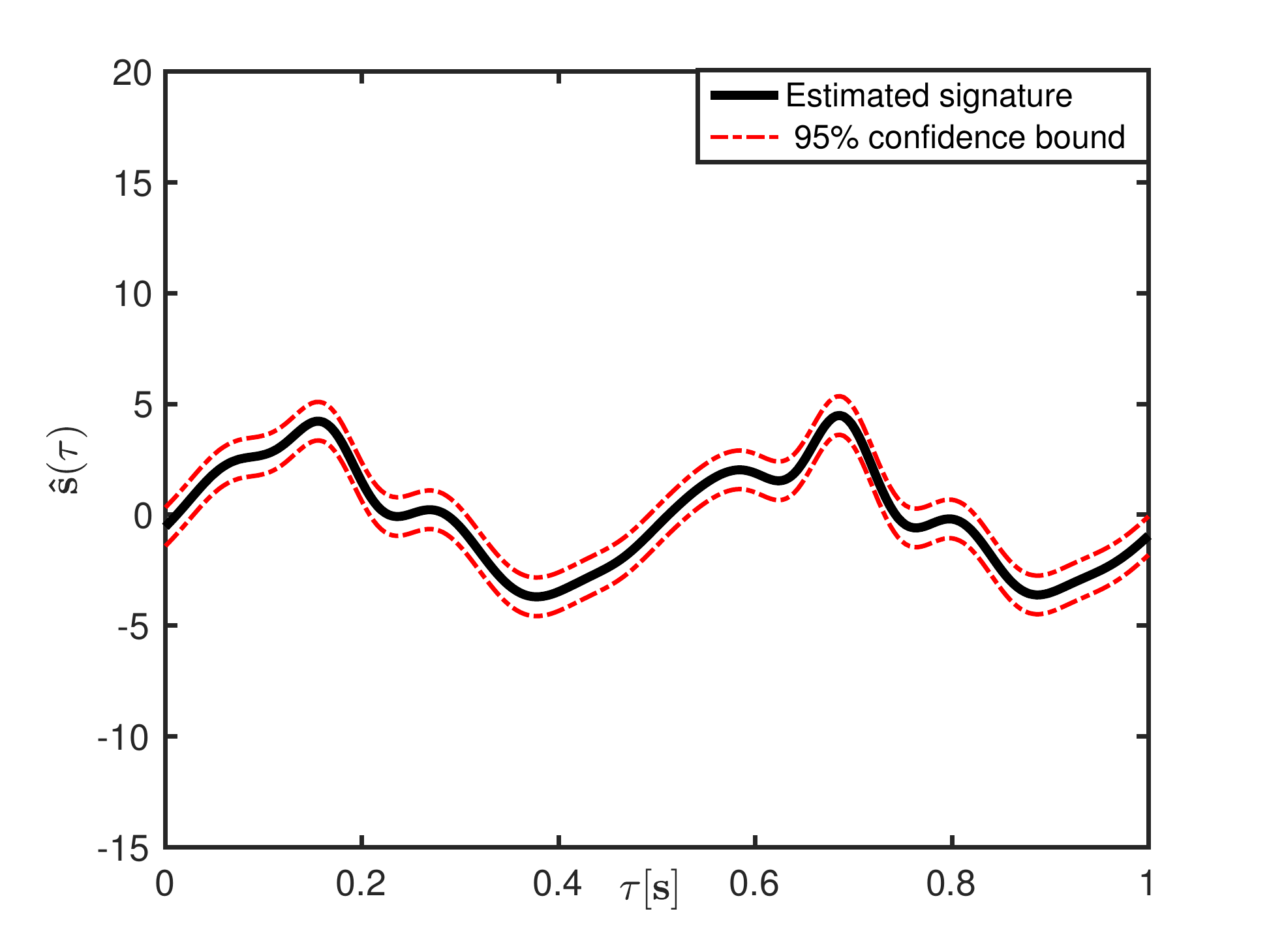}}
		\subfloat[ R1 \label{RFHshat}]{\includegraphics[width=0.35\textwidth]{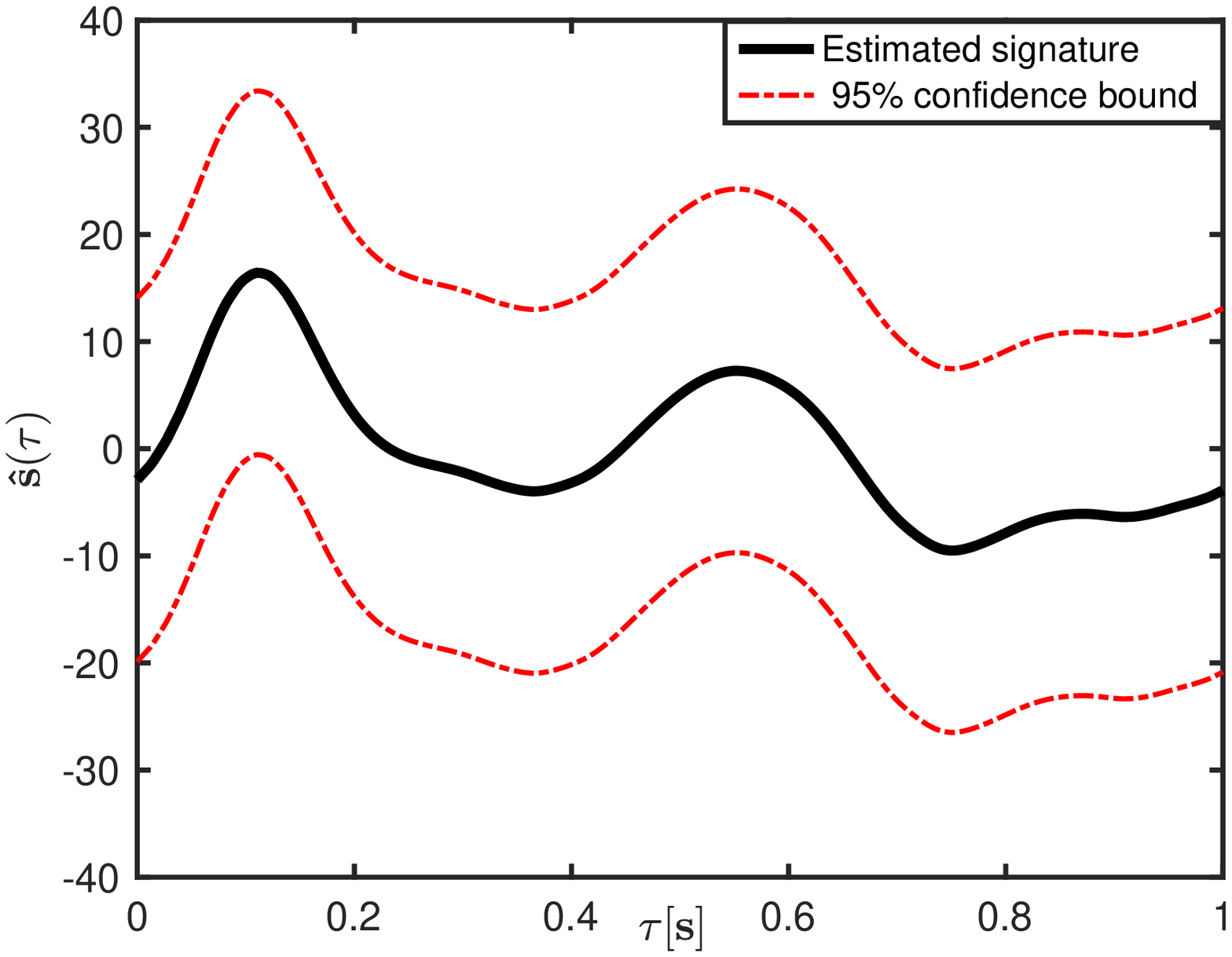}}\\
		\subfloat[ W2 \label{WSHshat}]{\includegraphics[width=0.35\textwidth]{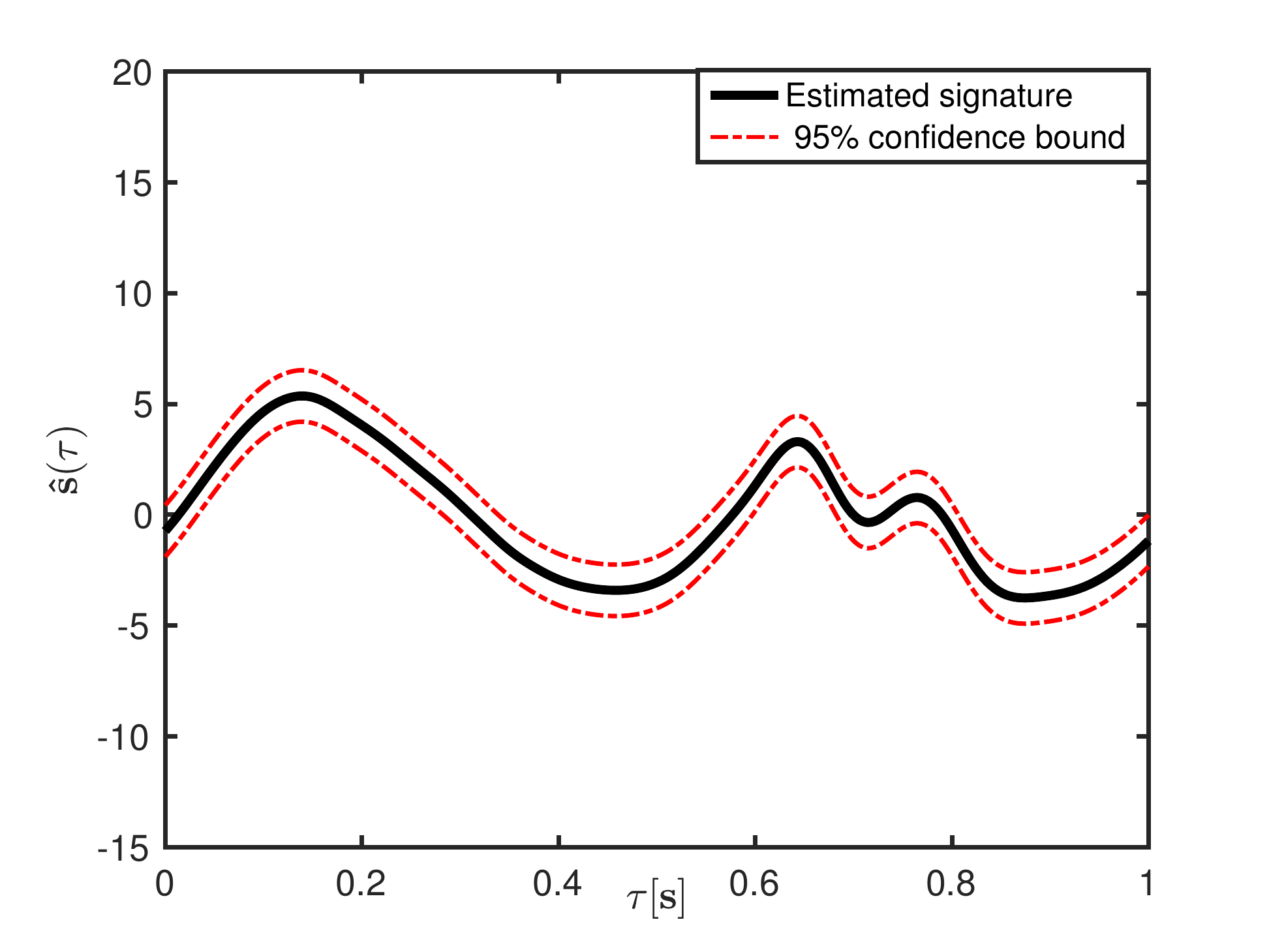}}
		\subfloat[ R2 \label{RSHshat}]{\includegraphics[width=0.35\textwidth]{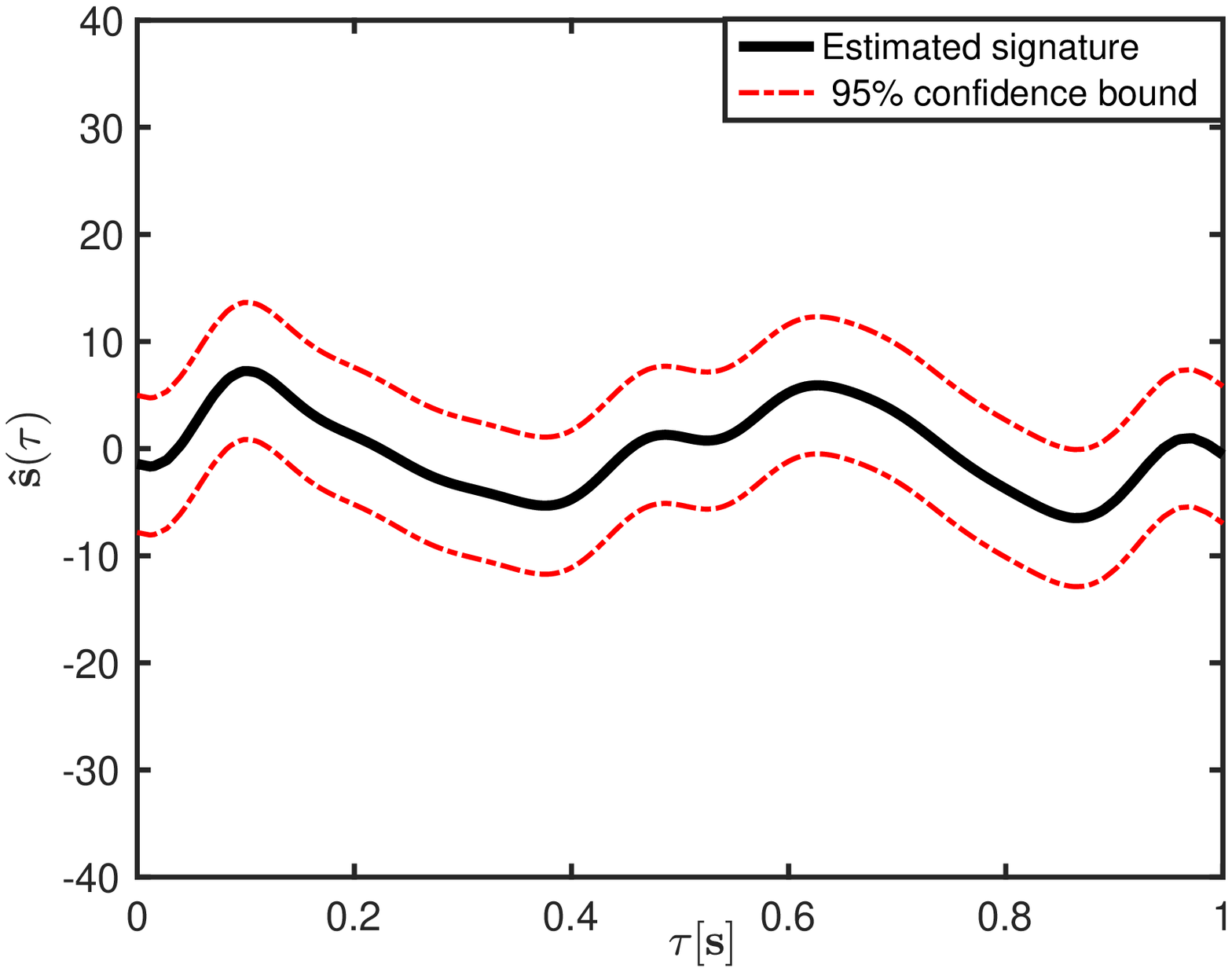}}\\
		\subfloat[ W3 \label{WFPshat}]{\includegraphics[width=0.35\textwidth]{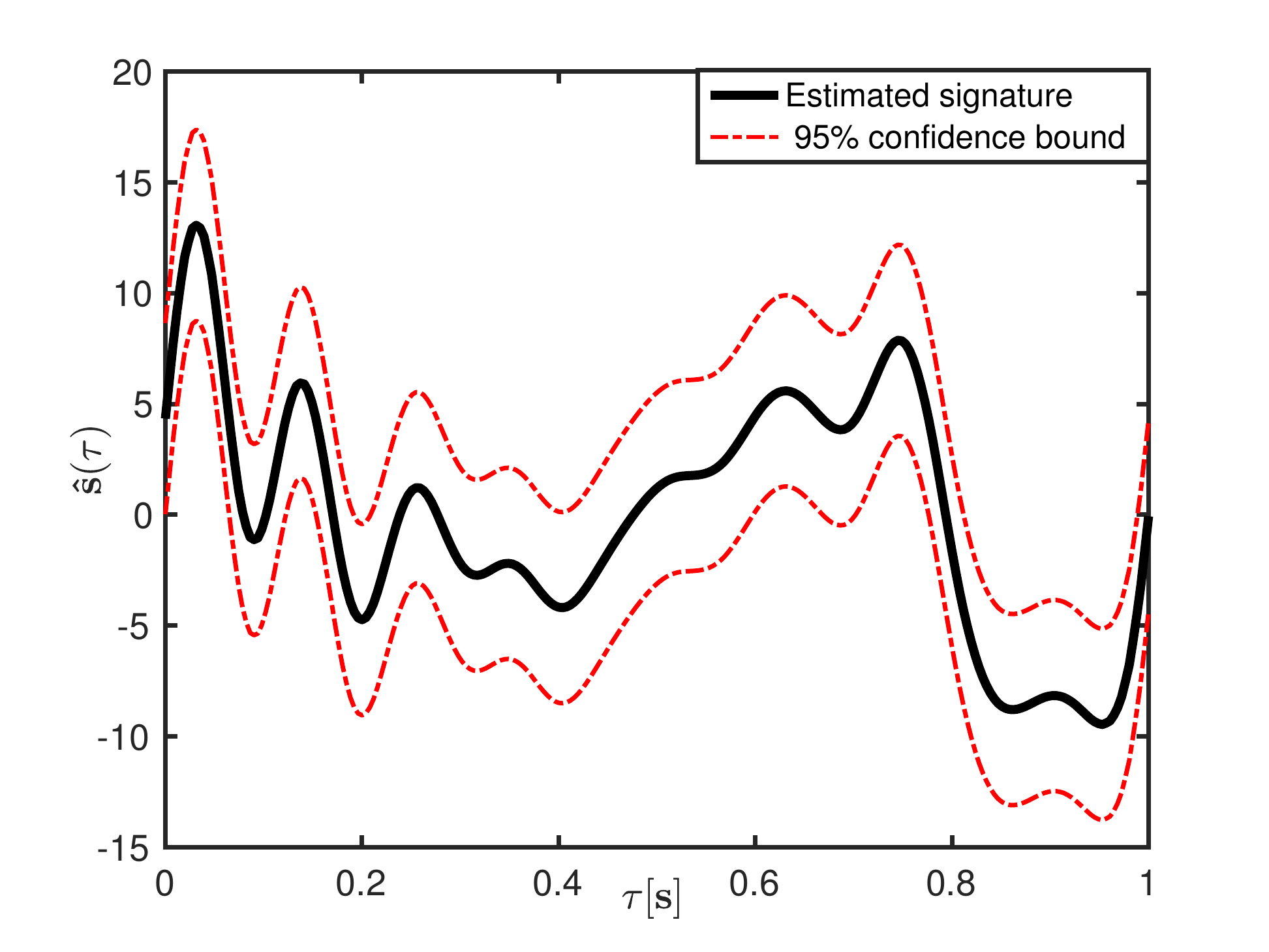}}
		\subfloat[ R3 \label{RFPshat}]{\includegraphics[width=0.35\textwidth]{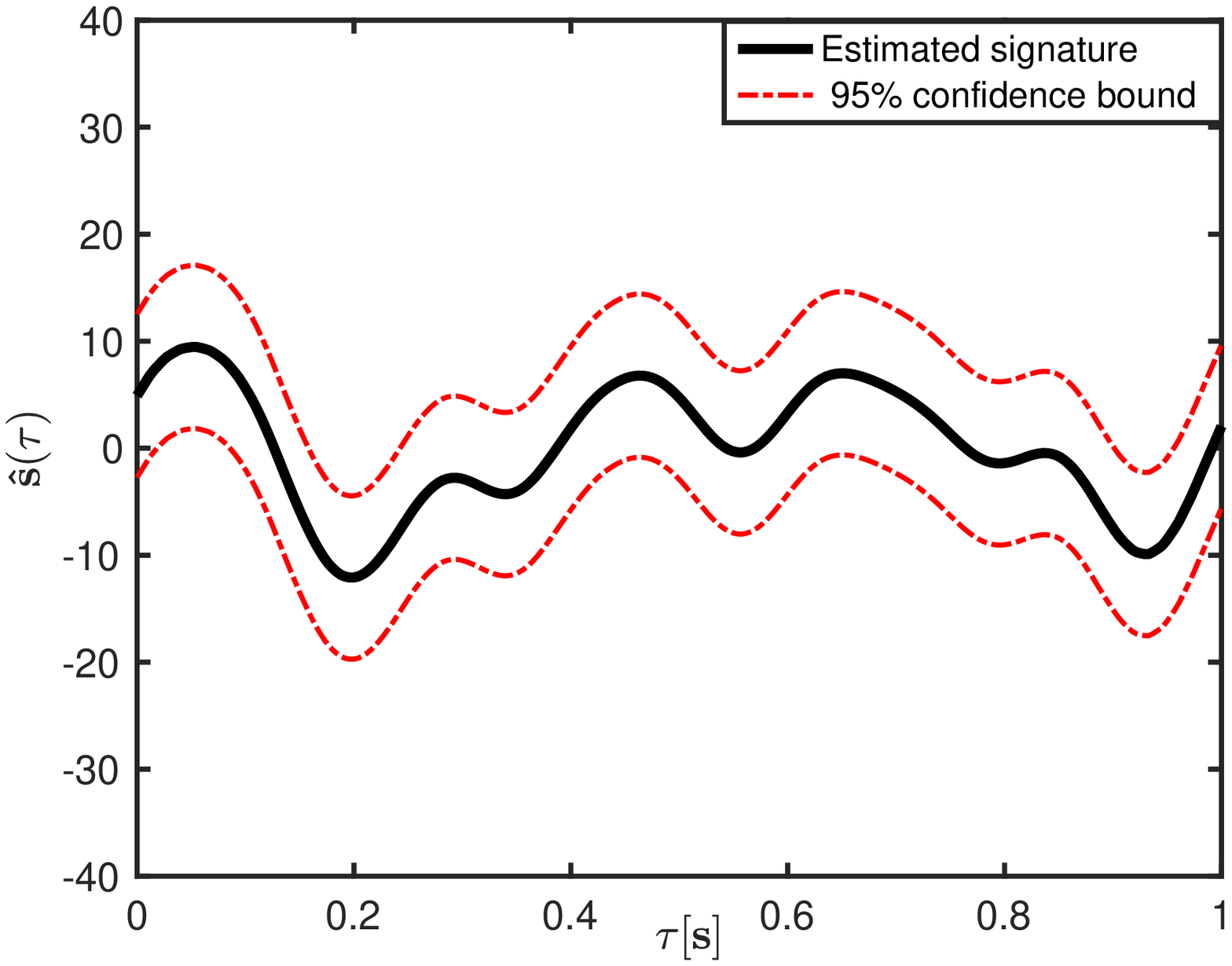}}\\
		\subfloat[ W4 \label{WBPshat}]{\includegraphics[width=0.35\textwidth]{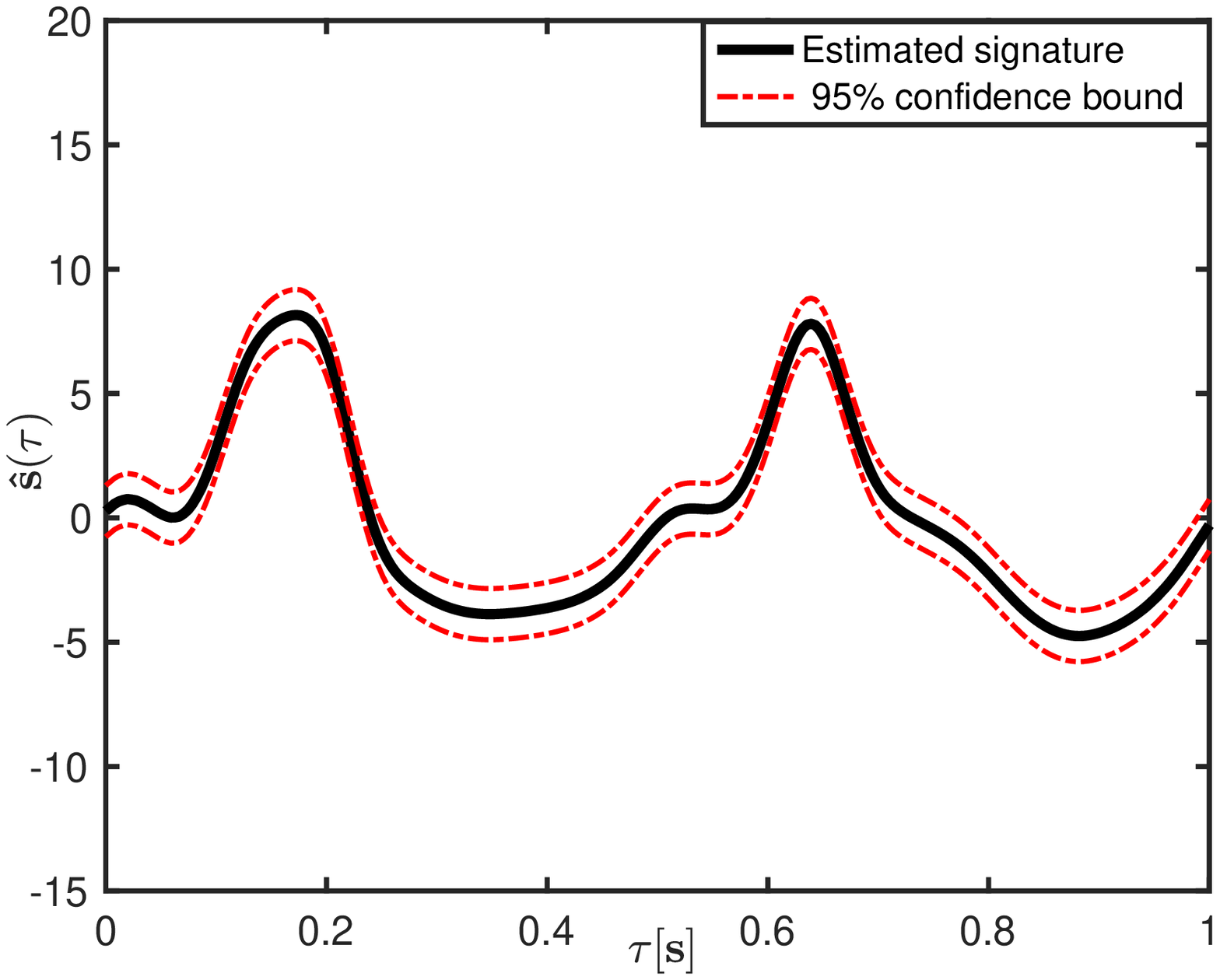}} 
		\subfloat[ R4 \label{RBPshat}]{\includegraphics[width=0.35\textwidth]{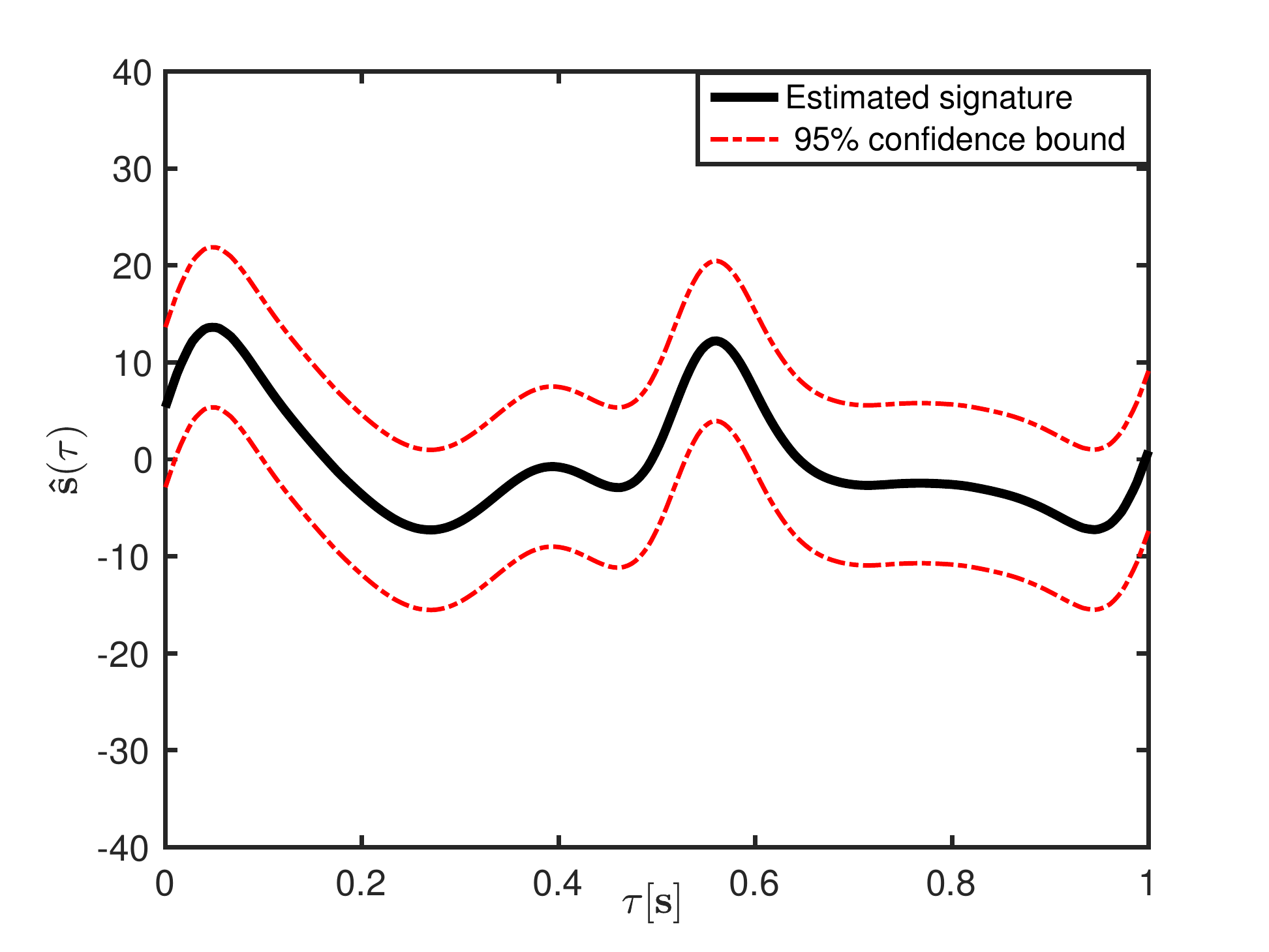}}		
	\end{center}
	\caption{Gait signatures for all eight scenarios introduced in Table~\ref{tbl:classes} [top] walking, [bottom] running.}
	\label{shat_modes}
\end{figure}
\begin{figure}[!t] 
	\begin{center}
		\subfloat[Walking\label{EnergyWalk}]{\includegraphics[width=0.5\columnwidth]{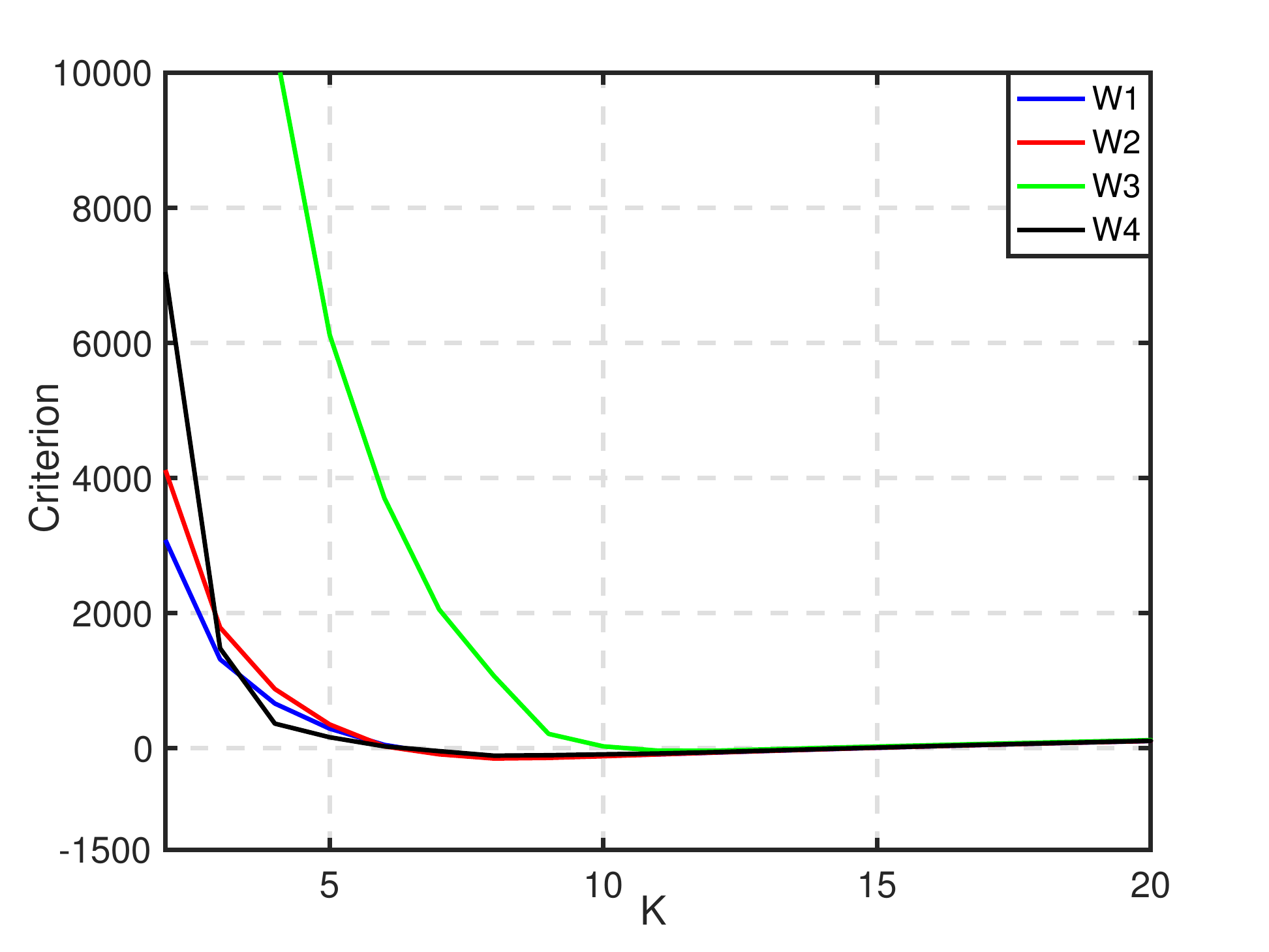}}
		\subfloat[Running\label{REnergyUN}]{\includegraphics[width=0.5\columnwidth]{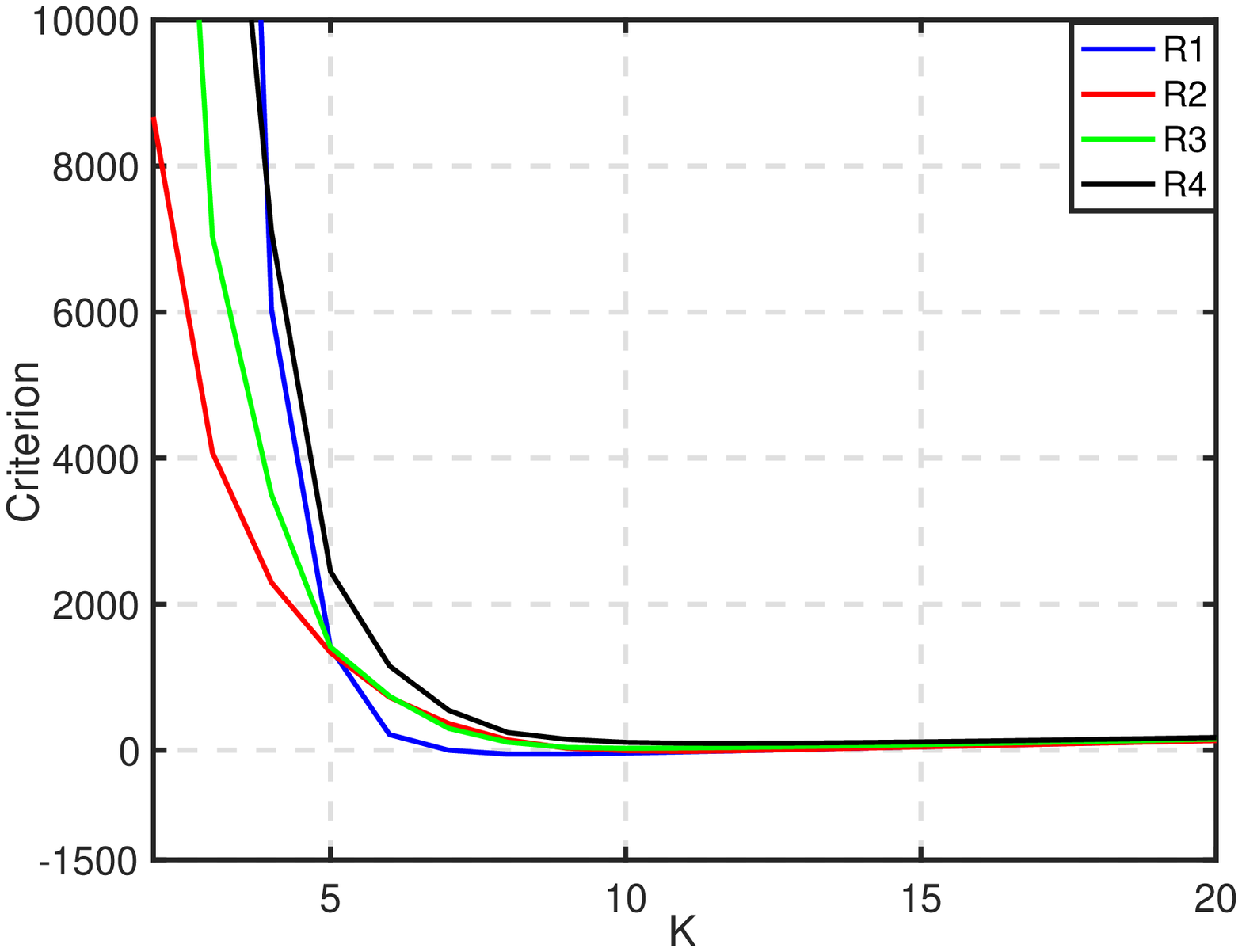}}
	\end{center}
	\caption{Model order selection using BIC.}
	\label{MOall}
\end{figure}
The updated gait signatures for all eight scenarios in Table~\ref{tbl:classes}, obtained from the tuned gait cycles, together with their corresponding $95\%$ confidence bounds, are presented in Fig.~\ref{shat_modes}. As the figures suggest, the estimated gait signatures for all considered scenarios have a unique pattern. For walking modes, all signatures have a  very narrow confidence bound which is a verification that the tuned gait cycles resemble the final gait signatures.  The bound for running mode, however, is wider especially for the fixed  hand device modes. This can also be verified by  Fig.~\ref{Rbox}, in which the variance of all optimal gait cycles for ``R1'' is higher than the other scenarios. This uncertainty can be explained by noting that users might have various unpredictable hand movements (especially while running) compared to the backpack and pocket scenarios in which the phone is more or less in a fixed position.

The updated signatures could be used for further investigation such as extracting low-dimensional feature vectors  for the gait cycles. Twenty different model orders are applied to all introduced scenarios to find the best model order for each of them. The Bayesian information criterion for all $20$ model orders  are presented in Fig.~\ref{MOall}. As shown in the figures, considering $K=8$ for walking and $K=10$ for running case would be suitable model orders to extract  the feature vectors for the gait cycle in most cases, hence the signatures can be encoded efficiently.  However, pocket mode in walking motion  is the worst case from an FS perspective and requires $K=10$. 

Finally, the uniqueness of the signatures  can be verified by investigating the scalar product of of a pair of signatures. By doing so, it can be seen that the most unique signature corresponds to ``W3'' and  that “R3” has the least similarity to the others. However, this is not always the case and there are cases that are not well separable from each other. For example, the signatures obtained for ``W1'', ``W4'' and ``R2'' have higher correlation to each other.
\newcommand\items{8}   
\arrayrulecolor{white} 
\begin{table}
	\caption{Correlation matrix for different signatures.}
	\centering
	\scalebox{1}	{
		{\noindent\begin{tabular}{cc*{\items}{|E}|}
				\multicolumn{1}{c}{} &\multicolumn{1}{c}{} &\multicolumn{\items}{c}{Classes} \\ \hhline{~*\items{|-}|}
				\multicolumn{1}{c}{} & 
				\multicolumn{1}{c}{} & 
				\multicolumn{1}{c}{\rotatebox{45}{W1}} & 
				\multicolumn{1}{c}{\rotatebox{45}{W2}} & 
				\multicolumn{1}{c}{\rotatebox{45}{W3}} &
				\multicolumn{1}{c}{\rotatebox{45}{W4}} & 
				\multicolumn{1}{c}{\rotatebox{45}{R1}}&
				\multicolumn{1}{c}{\rotatebox{45}{R2}}&
				\multicolumn{1}{c}{\rotatebox{45}{R3}}&
				\multicolumn{1}{c}{\rotatebox{45}{R4}} \\ \hhline{~*\items{|-}|}
				\multirow{\items}{*}{\rotatebox{90}{Classes}} 
				&W1  &1    &0.81    &0.46    &0.84    &0.58    &0.82    &0.26    &0.47     \\ \hhline{~*\items{|-}|}
				&W2  &0.81    &1    &0.32    &0.82   &0.59    &0.66   &0.01    &0.29     \\ \hhline{~*\items{|-}|}
				&W3  &0.46    &0.32    &1   &0.43   & 0.32   & 0.44    &0.39    &0.39     \\ \hhline{~*\items{|-}|}
				&W4  & 0.84    &0.82  &0.43  &1  &0.60    &0.80    &0.09    &0.35     \\ \hhline{~*\items{|-}|}
				&R1  &0.5    &0.59    &0.32    &0.60    &1    &0.68    &0.19    &0.66    \\ \hhline{~*\items{|-}|}
				&R2  &0.82  &0.66    &0.44    &0.80    &0.68    &1    &0.37    &0.50     \\ \hhline{~*\items{|-}|}
				&R3  & 0.26   &0.01    &0.39    &0.09    &0.19    &0.37    &1    &0.57   \\ \hhline{~*\items{|-}|}
				&R4  &0.47    &0.29    &0.39    &0.35    &0.66    &0.50    &0.57    &1   \\ \hhline{~*\items{|-}|}
				\label{tbl:corr}
		\end{tabular}} 
	}
\end{table}

In order to make the algorithm more automated, classification of the scenarios introduced in Table~\ref{tbl:classes}  would be an interesting topic for further investigation. For example, it is possible to consider the correlation of the  gait signatures, given in Table~\ref{tbl:corr}, as inputs to the classifier. As a low hanging fruit, consider a classifier based on the highest correlation score for which we achieved~70\%  classification accuracy. However, more sophisticated classifiers, {\em e.g.} those that consider temporal correlations in a filtering framework, would achieve much higher classification accuracies of 90-98\%, see~\cite{Journal:Kasebzadeh2019-2} for more detail.  
%
\section{Conclusion}
\label{sec:con}
Reliable pedestrian navigation systems require accurate step length estimation which in turn requires accurate gait cycle detection. In this work, an algorithm has been proposed for accurate gait cycle segmentation using IMU signals in multiple device and motion mode scenarios. For this purpose, we first used a classical thresholding algorithm to detect the gait cycles. Then, based on the asynchronous averaging of the gait cycles, a unique signature for each scenario was estimated. Furthermore, as a post-processing step, an optimization-based solution was proposed to tuned  the segmentation of the IMU signals in a way  that  minimized the variance of signature for each gait cycle.  We showed that a Fourier series expansion of gait signatures provides a low-dimensional feature vector which could possibly be highly beneficial together with the final least square cost value and the step time variations for  classification purposes. The performance of the proposed method has been evaluated using  measurements collected from IMUs embedded in smartphones for different motion modes while being carried  in different device modes. 
The results indicate good performance for the gait cycle segmentation problem for all of the considered scenarios.

\section{Acknowledgments}
The author would like to thank  PhD students from Link\"oping University who voluntarily participated in the data collection experiment.

This work is funded by the European Union FP7, the Marie Curie training program on
 {\em Tracking in Complex Sensor Systems} (TRAX) with grant number 607400,  and the Swedish Research Council project {\em Scalable Kalman Filter}. 
\bibliographystyle{unsrtnat}
\bibliography{J1_v8_arxiv}

\end{document}